\documentclass[11pt]{article}
\usepackage{amsmath,amssymb,enumitem,mathtools}
\usepackage{algpseudocode,algorithm}
\usepackage{xcolor}
\usepackage{subfig}
\usepackage{microtype}
\usepackage{graphicx}
\graphicspath{{./figures/}}

\usepackage{epstopdf}
\usepackage{array}

\usepackage{booktabs}
\usepackage{fullpage}
\usepackage{cite}
\newcolumntype{L}{>{\centering\arraybackslash}m{4cm}}
\newcommand{\specialcell}[2][c]{%
  \begin{tabular}[#1]{@{}c@{}}#2\end{tabular}}

\algrenewcommand\algorithmicindent{2em}

\algblockdefx{MRepeat}{EndRepeat}{\textbf{repeat}}{}
\algnotext{EndRepeat}

\newcounter{proxy}[table]
\newcommand{\subjto}{\text{subject to}}
\newcommand{\prox}{\text{prox}}
\DeclareMathOperator*{\mmz}{\text{minimize}}
\DeclareMathOperator*{\argmin}{arg\,min}

\begin{document}

\title{Fraction-variant beam orientation optimization for non-coplanar IMRT} %Title of paper

\author{Daniel O'Connor \thanks{Department of Radiation Oncology, University of California Los Angeles}
\and
Dan Nguyen \thanks{Department of Radiation Oncology, UT Southwestern}
\and
Dan Ruan \footnotemark[1]
\and
Victoria Yu \footnotemark[1]
\and
Ke Sheng \footnotemark[1]}

%\date{\today}
\date{}
\maketitle

\begin{abstract}
Conventional beam orientation optimization (BOO) algorithms for IMRT assume that the same set of beam angles is used for all treatment fractions. 
In this paper we present a BOO formulation based on group sparsity that simultaneously optimizes non-coplanar beam angles for all fractions, yielding a fraction-variant (FV) treatment plan.
Beam angles are selected by solving a multi-fraction fluence map optimization problem involving 500-700 candidate beams per fraction, with an additional group sparsity term that encourages most candidate beams to be inactive. 
The optimization problem is solved using 
the Fast Iterative Shrinkage-Thresholding Algorithm. 
Our FV~BOO algorithm is used to create non-coplanar, five-fraction treatment plans for prostate and lung cases, as well as a non-coplanar 30-fraction plan for a head and neck case. 
A homogeneous PTV dose coverage is maintained in all fractions.
The treatment plans are compared with fraction-invariant plans that use a fixed set of beam angles for all fractions.
The FV plans reduced mean and max OAR dose on average by 3.3\% and 3.7\% of the prescription dose, respectively.
Notably, mean OAR dose was reduced by 14.3\% of prescription dose (rectum), 11.6\% (penile bulb), 10.7\% (seminal vesicle), 5.5\% (right femur),
3.5\% (bladder), 4.0\% (normal left lung), 
15.5\% (cochleas), and 5.2\% (chiasm).
Max OAR dose was reduced by 14.9\% of prescription dose (right femur), 8.2\% (penile bulb), 12.7\% (proximal bronchus),
4.1\% (normal left lung), 15.2\% (cochleas), 10.1\% (orbits), 9.1\% (chiasm), 8.7\% (brainstem), and 7.1\% (parotids).
Meanwhile, PTV homogeneity defined as $D_{95}$/$D_5$ improved from .95 to .98 (prostate case) and from .94 to .97 (lung case),
and remained constant for the head and neck case.
Moreover, the FV plans are dosimetrically similar to conventional plans that use twice as many beams per fraction.
Thus, FV~BOO offers the potential to reduce delivery time for non-coplanar IMRT.

%The FV~BOO plans improve dosimetry without increasing the number of beams per fraction.
%Alternatively, FV~BOO plans are able to maintain similar dosimetry while using only half as many beams per fraction,
%reducing the delivery time for non-coplanar IMRT. 
\end{abstract}


\section{Introduction}

Compared to coplanar IMRT and volumetric modulated arc therapy (VMAT),
non-coplanar IMRT with beam orientation optimization (BOO) has been shown to improve dose compactness 
and organ-at-risk (OAR) sparing substantially for various cancer sites~\cite{woods2016viability, dong20134pi, dong20134pi_2,dong2014feasibility, nguyen2014feasibility, nguyen2014integral,rwigema20154pi, dong2014feasibility_2}.
However, a large number of non-coplanar beams (typically 15 or more) are needed to maximize dosimetric improvement,
resulting in a longer treatment time that may not be clinically viable.
This problem is compounded by the additional maneuvering of the gantry and couch
that is needed for non-coplanar delivery. 
To overcome this challenge, we examine a traditional but seemingly unnecessary constraint in radiotherapy planning --- namely, the use of a fixed set 
of beam angles for all fractions of the treatment course. 
By removing this constraint and utilizing different beam angles in different fractions, 
the same dosimetric improvement may be achieved while using fewer beams per fraction, thus avoiding long treatment delivery times.

The dosimetric benefit of spatiotemporally non-uniform fractionation schemes
has been investigated recently in the papers~\cite{unkelbach2013simultaneous,unkelbach2015non,unkelbach2015emergence}, which 
solve a BED-based multi-fraction fluence map optimization problem
to obtain treatment plans that deliver distinct dose distributions in different fractions.
Spatiotemporally modulated radiotherapy has also been investigated in the papers~\cite{kim2009markov,kim2015feasibility,kim2016feasibility,saberian2017spatiotemporally},
which optimize fractionation schedules based on a linear-quadratic cell survival model.
%In these studies, the potentially superior tumor response to spatially and temporally varying
%However, in these studies, the potentially superior tumor response to spatially and temporally varying prescription doses was studied and 
However, in these studies, the treatment plans are 
optimized using the \emph{same} set of beams or VMAT arcs in all fractions,
and the idea of fraction-variant beam angle selection is not considered.
Moreover, the motivation of these studies is not to improve the cumulative physical dose distribution compared
to conventional planning methods but rather to improve the biological effectiveness, 
which may vary depending on the underlying tumor biology.
In the context of proton therapy, a modality known as single-field uniform dose (SFUD) delivers
uniform and constant dose to the tumor using varying beam angles.
Faced with a similar challenge of minimizing the time for gantry rotation in proton therapy, 
SFUD delivery alternates between several pre-selected beams and avoids high entrance dose to any particular volume~\cite{lomax1999intensity}.
In SFUD planning, each field is independently optimized to cover the target uniformly, while OAR doses vary between fields.
In addition to providing for a shorter treatment delivery time, 
SFUD has been shown to be more robust than multifield optimization intensity-modulated proton therapy (MFO-IMPT).
However, due to the more restrictive optimization space, 
SFUD dosimetry has been found inferior to that of MFO-IMPT~\cite{harding2014benchmarking,kirk2015comparison,zhu2011patient,stuschke2012potentials}.

%Despite these surface-level similarities,
Despite this related work,  
we are not aware of any previous fraction-variant beam orientation optimization studies.
In this paper we investigate the potential improvement in \emph{cumulative physical dose} distribution
from allowing beam orientations to vary between fractions. 
We present a BOO formulation based on group sparsity that simultaneously optimizes non-coplanar beam angles for all fractions, 
yielding fraction-variant treatment plans.

\section{Methods}
\label{methods}

\subsection{Notation}
Before presenting a problem formulation for fraction-variant beam orientation optimization,
we first establish some notation.
\begin{itemize}
\item $F$ is the total number of treatment fractions. 
We use $F = 30$ for conventional fractionation, or $F = 5$ for SBRT.
\item $B$ is the number of candidate beams \emph{per fraction}. 
Typically $B$ is between 500 and 700 for non-coplanar IMRT cases after colliding beams are removed.
\emph{We assume that the pool of candidate beam angles is the same for each fraction,
but we make no requirement that the same beam angles are used during each treatment session.}
The total number of candidate beams (including all candidate beams from all fractions)
is $BF$.
\item The vector $x_{f,b}$ is the fluence map for the $b$th candidate beam on treatment day $f$. 
\item 
The vector $x_f$
(for $f = 1,\ldots, F$)
is the concatenation of the vectors $x_{f,b}$, and the vector $x$ is the concatenation of the vectors $x_f$:
\[
x_f = \begin{bmatrix} x_{f,1} \\ x_{f,2} \\ \vdots \\ x_{f,B} \end{bmatrix} \qquad \text{and} \qquad
x = \begin{bmatrix} x_1 \\ x_2 \\ \vdots \\ x_F \end{bmatrix}.
\]
\item The matrix $A_i = \begin{bmatrix} A_i^1 & A_i^2 & \cdots & A_i^B \end{bmatrix}$ is the dose-calculation matrix for the PTV (when $i = 0$)
or for the $i$th OAR ($i = 1,\ldots,N$). Here $A_i^b$ is the block column
of $A_i$ corresponding to the $b$th beam firing position. 
Note that $A_i x_f = \sum_{b=1}^B A_i^b x_{f,b}$ is a vector
that records how much dose is delivered to each voxel of the $i$th structure
on treatment day $f$.
\item
For $i = 1,\ldots,N$, we define
\[
\bar A_i = \underbrace{\begin{bmatrix} A_i & A_i & \cdots & A_i \end{bmatrix}}_{F \text{ copies of } A_i},
\]
where the matrix $A_i$ is repeated $F$ times.
A key point is that only one copy of $A_i$ must be stored in computer memory, so introducing $\bar A_i$
does not increase RAM requirements.
Notice that the vector
\[
\bar A_i x = \sum_{f=1}^F A_i x_f
\]
stores the total dose delivered to each voxel in the $i$th OAR, summed over all treatment fractions.
\item 
The vector $d_0$ stores the prescription doses for each voxel in the PTV, 
and the vectors $d_i$ (for $i = 1, \ldots, N$) store prescribed maximum
doses for each voxel in the $i$th OAR.
\item The notation $y_+$ (where $y$ is a vector) is defined by $y_+ = \max(y,0)$, with the maximum taken componentwise.
\item 
The function $\| \cdot \|_1^{(\mu)}$ is the Huber penalty (with parameter $\mu > 0$), defined by
\begin{align}
\label{huberDef}
\| y \|_1^{(\mu)} &= \sum_j | y_j |^{(\mu)}, \\
\notag \qquad | y_j |^{(\mu)} &= 
\begin{cases} 
\frac{1}{2\mu} y_j^2 & \quad \text{if } |y_j| \leq \mu, \\
|y_j| - \frac{\mu}{2} & \quad \text{otherwise}.
\end{cases}
\end{align}
The notation $\| \cdot \|_1^{(\mu)}$ indicates that the Huber penalty is simply a smoothed out version
of the $\ell_1$-norm, and $\mu$ controls the amount of smoothing.
\item 
The matrix $D$ represents a discrete gradient operator, so that $D x$ is a list of intensity
differences between adjacent beamlets.
The function $\| D x \|_1^{(\mu)}$ is a smoothed total variation regularization penalty function.
Total variation regularization has been used in fluence map optimization to encourage
fluence maps to be piecewise constant, which enhances plan deliverability \cite{kim2012dose,zhu2008using,aapmPaper}.
\end{itemize}

\subsection{Fraction-variant beam orientation optimization}
\label{problemFormulationSection}
With the above notation in place, we are now ready to present a problem
formulation for fraction-variant (FV) beam orientation optimization.
We propose to select beam orientations for all fractions simultaneously by solving
a multi-fraction fluence map optimization problem involving a large number of candidate beams,
with an additional \emph{group sparsity} penalty term in the objective function that encourages most candidate
beams to be inactive. 
The group sparsity approach to beam orientation optimization is an established technique that was introduced in~\cite{jia2011beam}
and was shown to have state of the art performance in~\cite{aapm2016talk} and~\cite{liu2017new}.
In detail, our FV BOO problem formulation is 
\begin{align}
\label{fracVariantProb}
\mmz_x & \quad \underbrace{\sum_{f=1}^F \frac{\eta}{2} \| A_0 x_f - d_0/F  \|_2^2}_{\text{PTV}} 
+ \underbrace{\sum_{i=1}^N \frac{\alpha_i}{2} \| (\bar A_i x - d_i)_+ \|_2^2 + \frac{\beta_i}{2}\| \bar A_i x \|_2^2}_{\text{OARs}} \\
\notag &\quad + \underbrace{\gamma  \| D x \|_1^{(\mu)}}_{\text{fluence map deliverability}}
+ \underbrace{\sum_{f=1}^F \sum_{b=1}^B w_b \| x_{f,b} \|_2^{p}}_{\text{group sparsity}} \\
\notag \subjto & \quad x \geq 0,
\end{align}
where $0 < p < 1$.  We take $p = 1/2$ for all experiments in this paper. 
(Our motivation for choosing $p = 1/2$, as well as an explanation of why 
the value $p = 1$ is forbidden, is given in appendix~\ref{pAppendix}.)
The PTV terms encourage the PTV to be covered uniformly at each fraction,
so that $A_0 x_f \approx d_0/F$ for $f = 1,\ldots, F$.
The OAR terms $(\beta_i/2) \| \bar A_i x \|_2^2$ penalize
the total dose delivered to OARs (summed over all fractions), 
and the terms $(\alpha_i/2) \| (\bar A_i x - d_i)_+ \|_2^2$ specifically penalize
violations of the prescribed dose limits $ \bar A_i x \leq d_i$ for $i = 1,\ldots, N$.
The smoothed total variation regularization term $\gamma \| D x \|_1^{(\mu)}$
encourages all fluence maps for all fractions to be piecewise constant, for enhanced plan deliverability.
\emph{The group sparsity term encourages most candidate beams to be inactive}, 
while allowing for a different set of beams to be active at each fraction.
The remaining active (nonzero) beams for each fraction are the ones selected to be used during treatment.
In practice we find that approximately the same number of beams are selected for each fraction,
although this requirement is not enforced explicitly.
The weights $w_b$ can be chosen to be all the same, so that $w_b = c$ for all $b$, with the constant $c$ tuned
to control the number of active beams in the solution to~\eqref{fracVariantProb}.  Alternatively,
the weights $w_b$ can be computed (as a preprocessing step) according to the strategy explained in appendix~\ref{weightAppendix}.
Either way, the weights $w_b$ do not have to be tuned individually.
Once the beam angle selection step is complete, fluence maps for all selected beams
(for all fractions) are computed simultaneously by solving a multi-fraction fluence map optimization
problem which is similar to problem~\eqref{fracVariantProb}, but with no group sparsity term
and using only the beams which were selected for each fraction.

\subsection{Optimization algorithm}
\label{algorithmSection}

We solve the optimization problem~\eqref{fracVariantProb} using an accelerated proximal gradient method
known as the Fast Iterative Shrinkage-Thresholding Algorithm (FISTA)~\cite{beck2009FISTA,scheinberg2014fast}.
%\paragraph{Solution using FISTA.} 
FISTA solves optimization problems of the form
\begin{equation}
\label{fistaProb}
\mmz_{x \in \mathbb R^n} \, g(x) + h(x),
\end{equation}
where the convex function $g$ is assumed to be differentiable
(with a Lipschitz continuous gradient) and the function
$h$ is assumed to ``simple'' in the sense that its proximal operator
can be evaluated efficiently.  
The proximal operator (also known as ``prox-operator'') of $h$, with parameter $t > 0$, is defined by
\begin{equation}
\label{proxDef}
\prox_{th}(x) = \argmin_u \, h(u) + \frac{1}{2t} \|u - x \|_2^2.
\end{equation}
%(For the prox-operator of $h$ to be well defined, we require that $h$ is lower semi-continuous,
%which is a mild assumption that is usually satisfied in practice.)
%FISTA does not require $h$ to be differentiable.
Problem~\eqref{fracVariantProb} has the form \eqref{fistaProb},
where
\begin{align}
\label{gDef}
\notag g(x) &=
\sum_{f=1}^F \frac{\eta}{2} \| A_0 x_f - d_0/F  \|_2^2
+ \sum_{i=1}^N \left(\frac{\alpha_i}{2} \| (\bar A_i x - d_i)_+ \|_2^2 + \frac{\beta_i}{2}\| \bar A_i x \|_2^2 \right) + \gamma \| Dx \|_1^{(\mu)}
\end{align}
and 
\begin{equation}
\label{hDef}
h(x) = \begin{cases} \sum_{f=1}^F \sum_{b=1}^B w_b \| x_{f,b} \|_2^{p}  & \quad \text{if } x \geq 0, \\
\infty & \quad \text{otherwise}.
\end{cases}
\end{equation}
The function $h$ enforces the constraint $x \geq 0$ by returning the value $\infty$
when this constraint is not satisfied.  
%(Enforcing hard constraints in this manner is a standard technique in mathematical optimization.)
Standard convergence results for FISTA assume that $h$ is convex,
but we have found
that FISTA converges to a good solution when $p = 1/2$, in which case $h$
is not convex.

The FISTA with line search algorithm is recorded in algorithm~\ref{fistaAlg_lineSearch}.
\begin{figure}
\begin{algorithm}[H]
\caption{FISTA with line search}
\label{fistaAlg_lineSearch}
\begin{algorithmic}
\State Initialize $x_0$ and $t_0 > 0$, set $v_0 \coloneqq x_0$, select $0 < r < 1$, $s > 1$
\For{$k = 1,2,\ldots$}
\State $t \coloneqq s \, t_{k-1}$
\MRepeat
\State $\theta \coloneqq \begin{cases} 1 & \quad \text{if } k = 1 \\
\text{positive root of } t_{k-1} \theta^2 = t \theta_{k-1}^2(1 - \theta) & \quad \text{if } k > 1 \end{cases}$
\State $y \coloneqq (1 - \theta) x_{k-1} + \theta v_{k-1}$
\State $x \coloneqq \prox_{t h}(y - t \nabla g(y))$
\State \textbf{break if } $g(x) \leq g(y) + \langle \nabla g(y),x - y \rangle + \frac{1}{2t} \| x - y \|_2^2$
\State $t \coloneqq r t$
\EndRepeat
\State $t_k \coloneqq t$
\State $\theta_k \coloneqq \theta$
\State $x_k \coloneqq x$
\State $v_k \coloneqq x_{k-1} + \frac{1}{\theta_k}(x_k - x_{k-1})$
\EndFor
\end{algorithmic}
\end{algorithm}
\end{figure}
The key steps in each iteration of FISTA are to evaluate the gradient of
$g$ and the proximal operator of $h$.
To compute the gradient of $g$, we first note three facts that can be shown using basic calculus:
\begin{enumerate}
\item If $G(y) = \frac12 \| y \|_2^2$, then $\nabla G(y) = y$.
\item If $G(y) = \frac12 \| y_+ \|_2^2$, then $\nabla G(y) = y_+ = \max(y,0)$ (with maximum taken componentwise).
\item If $G$ is the Huber penalty function $G(y) = \| y \|_1^{(\mu)}$ (defined in equation~\eqref{huberDef}), then $\nabla G(y) = \frac{1}{\mu} P_{[-\mu,\mu]}(y)$,
where $P_{[-\mu,\mu]}(y)$ is the projection of the vector $y$ onto the set $\{ u \mid -\mu \leq u \leq \mu\}$.
(The inequalities are interpreted componentwise.)  Projecting onto this set is a simple componentwise ``clipping'' operation.
\end{enumerate}
It now follows from the chain rule that
\begin{align}
\label{gradgFormula}
\notag \nabla g(x) &=  \sum_{f=1}^F \eta A_0^T(A_0 x_f - d_0/F) + \sum_{i=1}^N \alpha_i \bar A_i^T(\bar A_i x - d_i)_+ + \beta_i \bar A_i^T \bar A_i x \\
& \quad + \frac{\gamma}{\mu} D^T P_{[-\mu,\mu]}(Dx).
\end{align}

To state a formula for the prox-operator of $h$, in the special case that $p = 1/2$,
we first express $h$ as $h(x) = \sum_{f=1}^F \sum_{b=1}^B h_{f,b}(x_{f,b})$,
where
\[
h_{f,b}(x_{f,b}) = \begin{cases} w_b \| x_{f,b} \|_2^p &  \quad \text{if } x_{f,b} \geq 0, \\
\infty & \quad \text{otherwise}.
\end{cases}
\]
(Recall that the vector $x_{f,b}$ is the fluence map for beam $b$ on treatment day $f$, and $x$ is the concatenation
of the vectors $x_{f,b}$.)
The vector $\prox_{th}(x)$ has the same size and block structure as $x$, 
and the separable sum rule for prox-operators~\cite{parikh2013proximal} informs us that the $(f,b)$th block of 
$\prox_{th}(x)$ is given by
\begin{equation}
\label{proxh}
\left[\prox_{t h}(x)\right]_{f,b}
=   \prox_{t h_{f,b}}(x_{f,b}).
\end{equation}
%In other words, to evaluate the $(f,b)$th block of the vector $\prox_{th}(x)$, we simply evaluate the prox-operator of the function $h_{f,b}$ at the point $x_{f,b}$.
We show in appendix~\ref{proxAppendix} that
\begin{equation}
\label{keyProx}
\prox_{t h_{f,b}}(x_{f,b}) = \prox_{t w_b \| \cdot \|_2^p}(\max(x_{f,b},0)).
\end{equation}
%Here $\prox_{t w_b \| \cdot \|_2^p}$ denotes the prox-operator of the function $\psi(x) = \|x \|_2^p$ with parameter $t w_b$. 
Here $\prox_{t w_b \| \cdot \|_2^p}$ denotes the prox-operator of the function $\|x \|_2^p$ with parameter $t w_b$. 
An explicit formula for this prox-operator in the special case that $p = 1/2$ is given in appendix~\ref{proxAppendix}.
Using formulas~\eqref{gradgFormula}, \eqref{proxh}, and~\eqref{keyProx} to compute the gradient of $g$ and the prox-operator
of $h$, it is now straightforward to solve problem \eqref{fracVariantProb} using the FISTA
with line search algorithm~\ref{fistaAlg_lineSearch}.

\subsection{Experimental setup}

\addtolength{\tabcolsep}{6pt}  
\begin{table}[htbp]
\begin{center}
\begin{tabular}{l c c}
\toprule
Case & Prescription dose (Gy) & PTV volume (cc) \\ \midrule
PRT & 40& 111.2  \\ % PRTAR
LNG & 48 & 72.3 \\ % LNGTM
H\&N   & 66 & 33.6  \\ % HNKLR
\bottomrule
\end{tabular}
\end{center}
\caption{Prescription dose and PTV volume for prostate, lung, and head and neck patients.}
\label{patientTable1}
\end{table}
\addtolength{\tabcolsep}{-6pt}

%\begin{table}[htbp]
%%\begin{ruledtabular}
%\begin{center}
%\begin{tabular}{l c L L c}
%\toprule
%Case & Num. fractions & Num. candidate beams per fraction & Avg. number of beams selected per fraction   & Runtime (hrs) \\ \midrule
%PRT & 5 & 674 &  9.6   & 1.5 \\ % PRTAR
%LNG & 5 & 520 &  6.6   & 1.1 \\ % LNGTM
%H\&N & 30 & 741 &  6.36   & 6.6  \\ % HNKLR
%\bottomrule
%\end{tabular}
%\end{center}
%%\end{ruledtabular}
%\caption{Number of fractions, number of candidate beams per fraction, average number
%of beams selected per fraction, and FISTA runtimes
%for cases ``PRT'', ``LNG'', and ``H\&N''.}
%\label{patientTable2}
%\end{table}

\begin{table}[htbp]
%\begin{ruledtabular}
\begin{center}
\begin{tabular}{l c c c c c}
\toprule
Case & \specialcell{Num.\\fractions} & \specialcell{Num. candidate \\ beams per fraction}& \specialcell{Total num.\\beamlets} & \specialcell{Avg. num. of beams\\ selected per fraction}   & \specialcell{Runtime\\(hrs)} \\ \midrule
PRT & 5 & 674 & 569,770 & 9.6  & 1.5 \\ % PRTAR
LNG & 5 & 520 & 347,240 & 6.6  & 1.1 \\ % LNGTM
H\&N & 30 & 741 & 2,013,870 & 6.36 & 6.6 \\ % HNKWL
\bottomrule
\end{tabular}
\end{center}
%\end{ruledtabular}
\caption{Number of fractions, number of candidate beams per fraction, 
total number of beamlets (for all candidate beams in all fractions),
average number of beams selected per fraction, and FISTA runtimes
for cases ``PRT'', ``LNG'', and ``H\&N''.}
\label{patientTable2}
\end{table}

%A prostate, a lung, and a head and neck patient were selected to test and evaluate the
An SBRT prostate, an SBRT lung, and a conventionally fractionated head and neck patient
with a single-level prescription dose to the unilateral lesion were selected to test and evaluate the proposed algorithm. 
The prescription doses and PTV volumes for each case are listed
in table~\ref{patientTable1}. For each case, we began with 1162 candidate beam firing positions evenly
distributed over the surface of a sphere, with approximately six degrees of separation between adjacent beams. A 3D human surface
measurement and a machine CAD model were utilized to identify collision zones,
%which of these 1162 candidate beam angles would result in collisions (such as collisions between gantry and patient, or between gantry and couch),
and beam angles that resulted in collisions were removed. 
%The details of collision space modeling were described previously~\cite{victoria2015development}.
As a result, 500-700 non-coplanar candidate beam firing positions, 
as reported in table~\ref{patientTable2}, were retained for dose calculation and optimization.
%(The exact number of candidate beams per fraction for each case is recorded in table~\ref{patientTable2}.)
Beamlet dose was calculated for all beams within the conformal aperture~+5~mm margin 
using convolution/superposition with a 6 MV polyenergetic kernel~\cite{neylon2014nonvoxel}. The dose calculation resolution
was isotropically 2.5~mm. The MLC leaf width at the isocenter was assumed to be 5~mm.

For each of the three cases, problem \eqref{fracVariantProb} was solved
using the FISTA with line search algorithm~\ref{fistaAlg_lineSearch}.
The weight parameters $\eta, \alpha_i, \beta_i$, and $\gamma$ appearing in the penalty functions
and the vectors $d_i$
%were tuned (on a case by case basis) by trial and error to achieve high quality treatment plans.
were tuned on a case by case basis to achieve high quality treatment plans.
%To reduce planning time, 
%These parameters were tuned with the number of fractions~$F$
%set equal to~$1$, in which case problem~\eqref{fracVariantProb} can be solved in about 5 minutes.
The Huber penalty smoothing parameter $\mu$ was set to $1$.
The beam weights $w_b$ were chosen as in equation~\eqref{weightFormula},
with the parameter $c$ in~\eqref{weightFormula}
tuned to yield a reasonable number of active beams per fraction.
The optimization variable $x$ was initialized randomly with components between $0$ and $1$,
drawn independently from a uniform distribution. 
The parameters $r$ and $s$ in algorithm~\ref{fistaAlg_lineSearch}
were taken to be $r = .5$ and $s = 1.25$. 
In each case FISTA was run for 5000 iterations, 
which was sufficiently many for the algorithm to converge on a stable selection of beams for each fraction.
The FISTA runtimes are shown in table~\ref{patientTable2}.
%A $2$-$3$ cm shell structure surrounding the PTV was utilized to penalize dose spillage to normal tissue. 

The resulting FV treatment plans were compared with treatment plans
that are conventional in the sense that
they use a fixed set of beam angles and fluence maps for all fractions.
The conventional plans were computed by taking $F = 1$ in problem~\eqref{fracVariantProb}.
(These conventional plans will sometimes be described as \textbf{fraction-invariant (FI)} to emphasize
that they are not FV.)
For each case, the conventional plans were computed using identical OAR weights as were used for the fraction-variant plan,
except that for case ``H\&N'' the skin structure weighting was increased when computing
the FI plans to prevent unacceptable hot spots on the skin.

For plan comparison, PTV $D_{98}$, $D_{99}$, and PTV homogeneity defined as $D_{95}$/$D_5$ were evaluated.
All treatment plans were scaled so that PTV $D_{95}$ was equal to the prescription dose.
OAR max and mean dose, denoted by $\text{D}_{\text{max}}^{\text{FV}}$ and $\text{D}_{\text{mean}}^{\text{FV}}$ 
for the fraction-variant plans
 and $\text{D}_{\text{max}}^{\text{FI}}$ and $\text{D}_{\text{mean}}^{\text{FI}}$ for the fraction-invariant plans, 
were also calculated for assessment.  
For each OAR, the difference in max dose and the difference in mean dose between the two plans were computed. 
Max dose is defined as the dose at 2 percent of the structure volume, $D_2$, which is recommended by the ICRU-83 report \cite{gregoire2011state}.

All treatment plans were created on a computer
with two Intel Xeon CPU~E5-2687W~v3 3.10~GHz processors and 512 GB of RAM.
%(This amount of RAM is not needed; the dose-calculation matrix typically
%requires about 8 GB of RAM, after downsampling, in our experiments.) 

\section{Results}
\label{numericalResults}

\begin{figure}[!htb]
\centering
\subfloat[\label{prt_allFracs}]{\includegraphics[width=.3\textwidth]{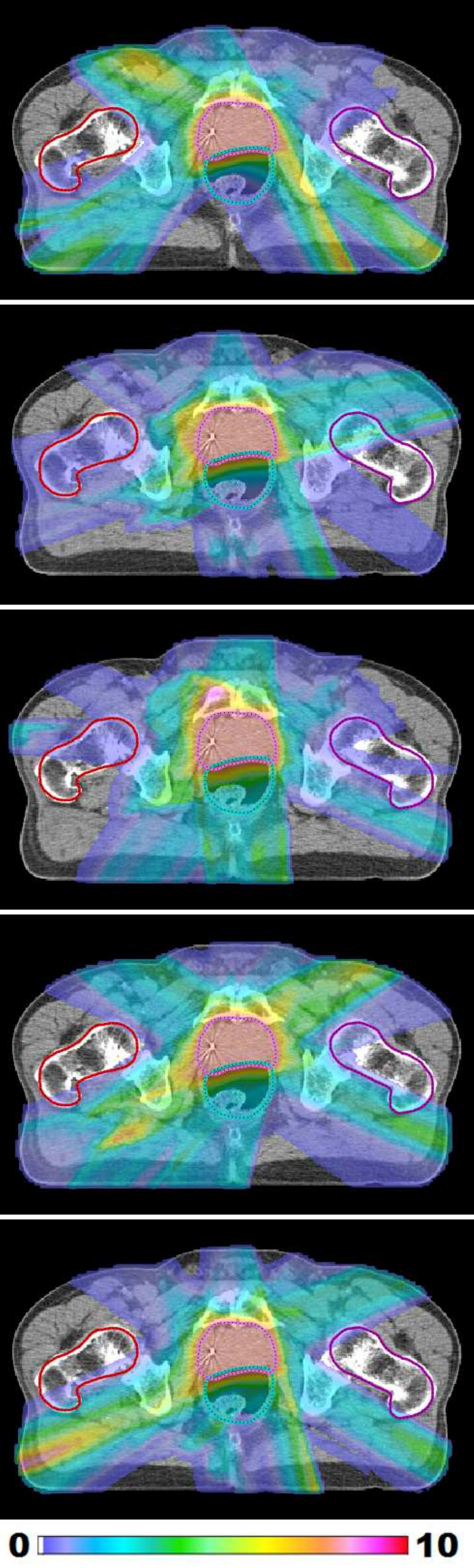}}\hfill
\subfloat[\label{lng_allFracs}]{\includegraphics[width=.3\textwidth]{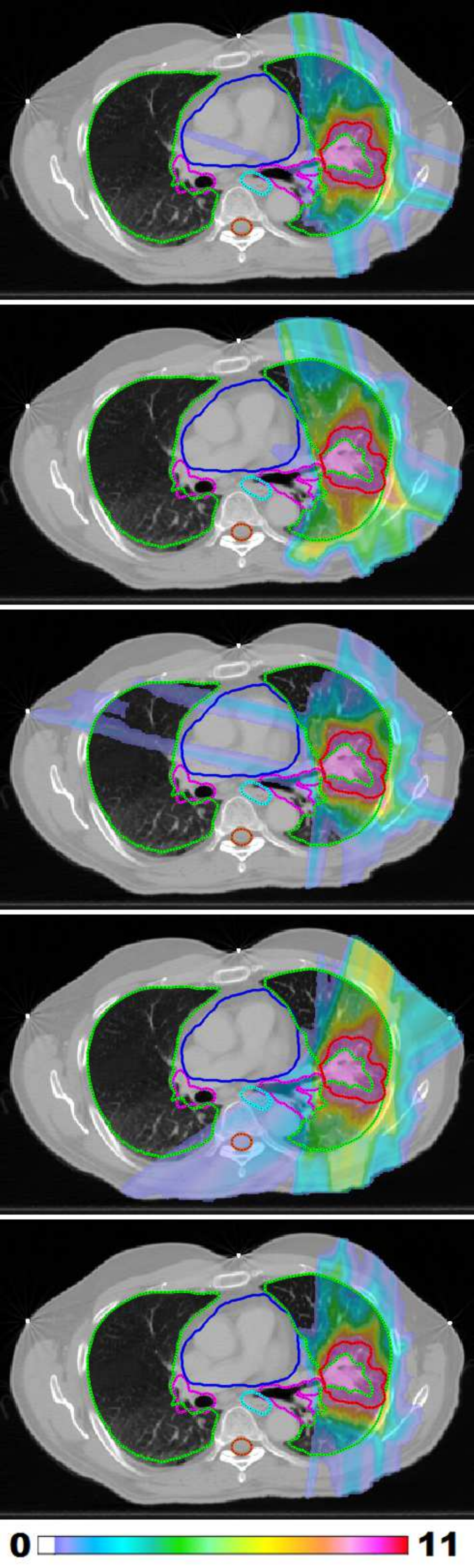}}\hfill
\subfloat[\label{hnk_allFracs}]{\includegraphics[width=.3\textwidth]{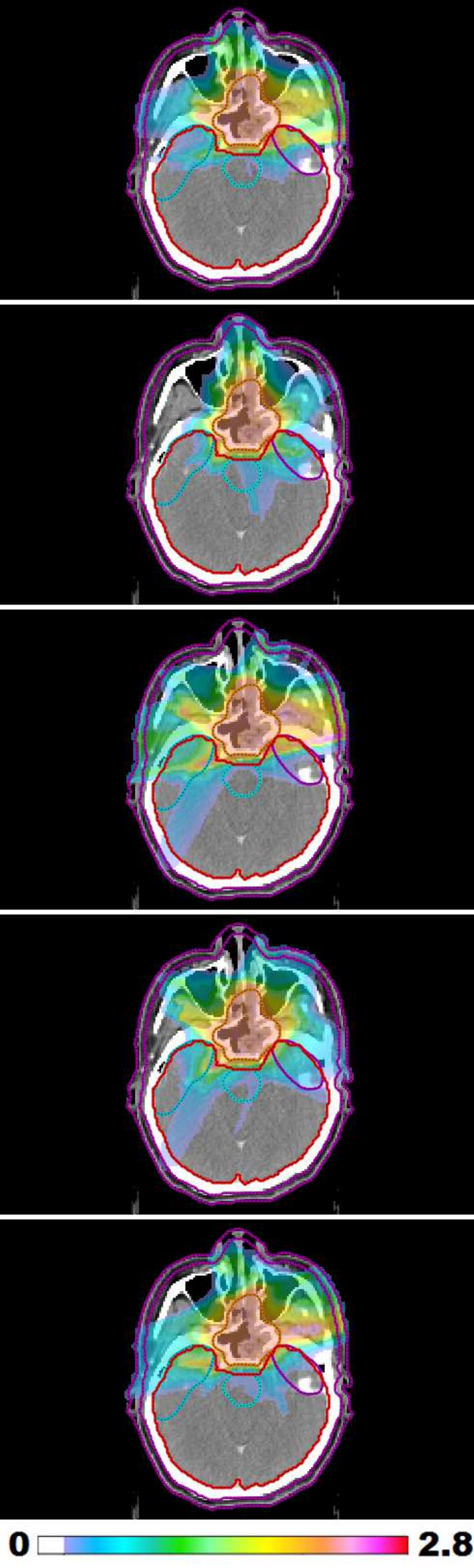}}
\caption{\protect\subref{prt_allFracs} and \protect\subref{lng_allFracs} 
Fraction-variant treatment plans for each of five fractions for cases ``PRT'' and ``LNG''. The PTV receives a uniform dose of $8$ Gy (approximately)
at each fraction for case ``PRT'' and 9.6 Gy (approximately) for case ``LNG''. 
\protect\subref{hnk_allFracs} Dose colormaps for fractions 1, 7, 13, 19, and 25 for case ``H\&N''. The PTV receives a uniform dose
of 2.2 Gy (approximately) throughout the treatment course.}
\label{allFracsStacked}
\end{figure}

\subsection{Prostate case}
\begin{figure}[!htb]
\centering
\subfloat[Fraction 1\label{prt_beams_frac1}]{\includegraphics[width=.3\textwidth]{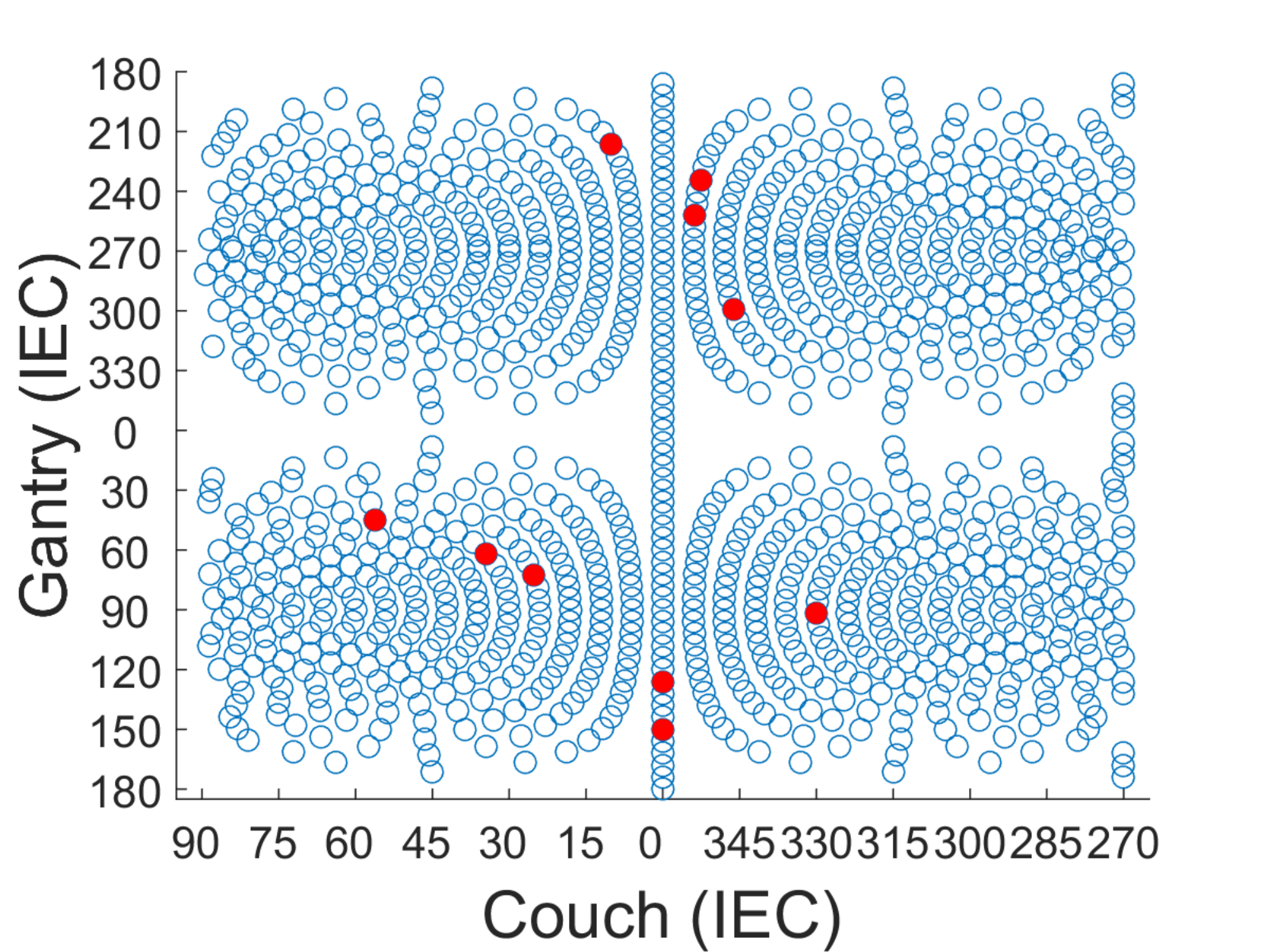}}\hfill
\subfloat[Fraction 2\label{prt_beams_frac2}]{\includegraphics[width=.3\textwidth]{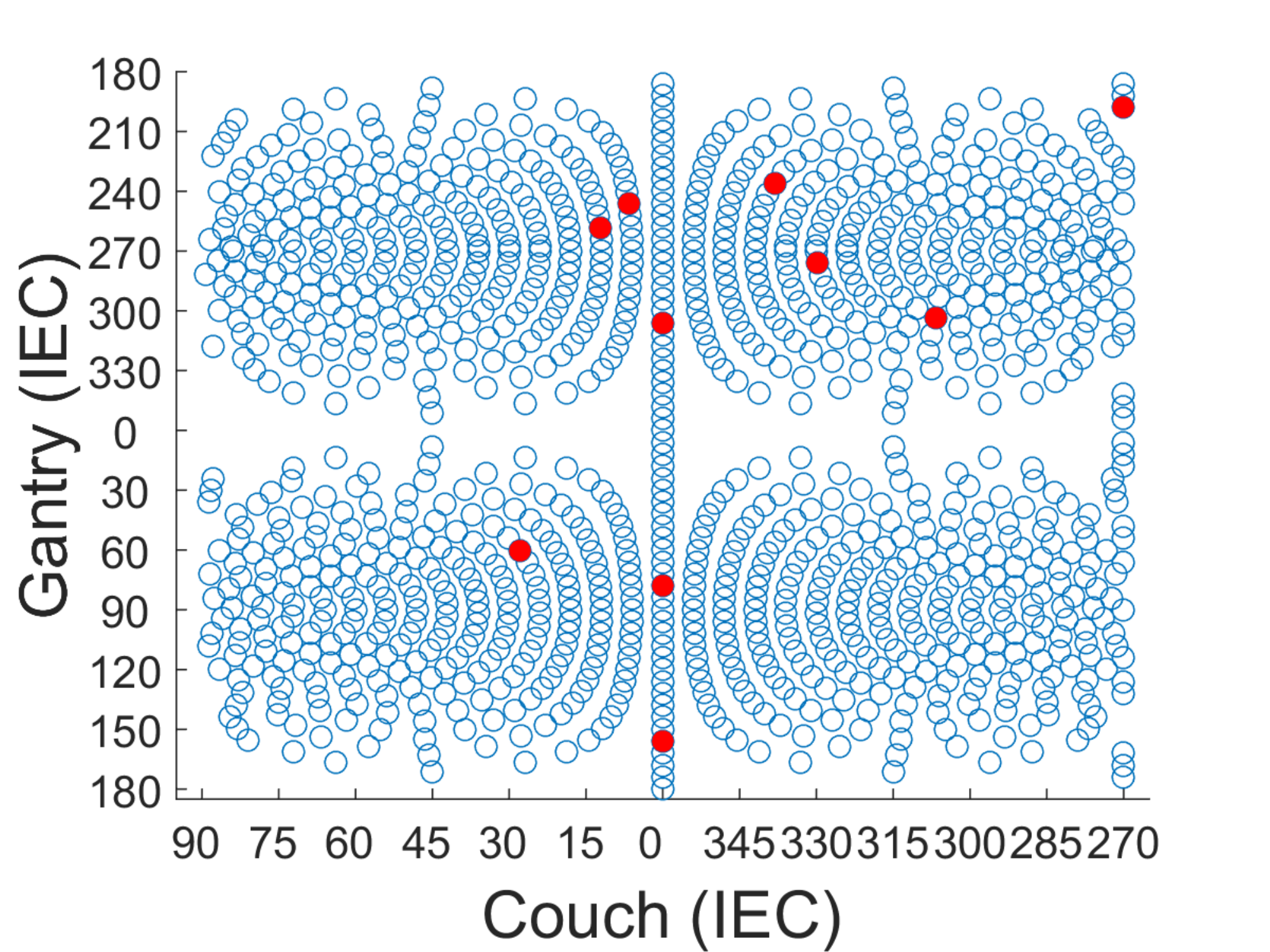}}\hfill
\subfloat[Fraction 3\label{prt_beams_frac3}]{\includegraphics[width=.3\textwidth]{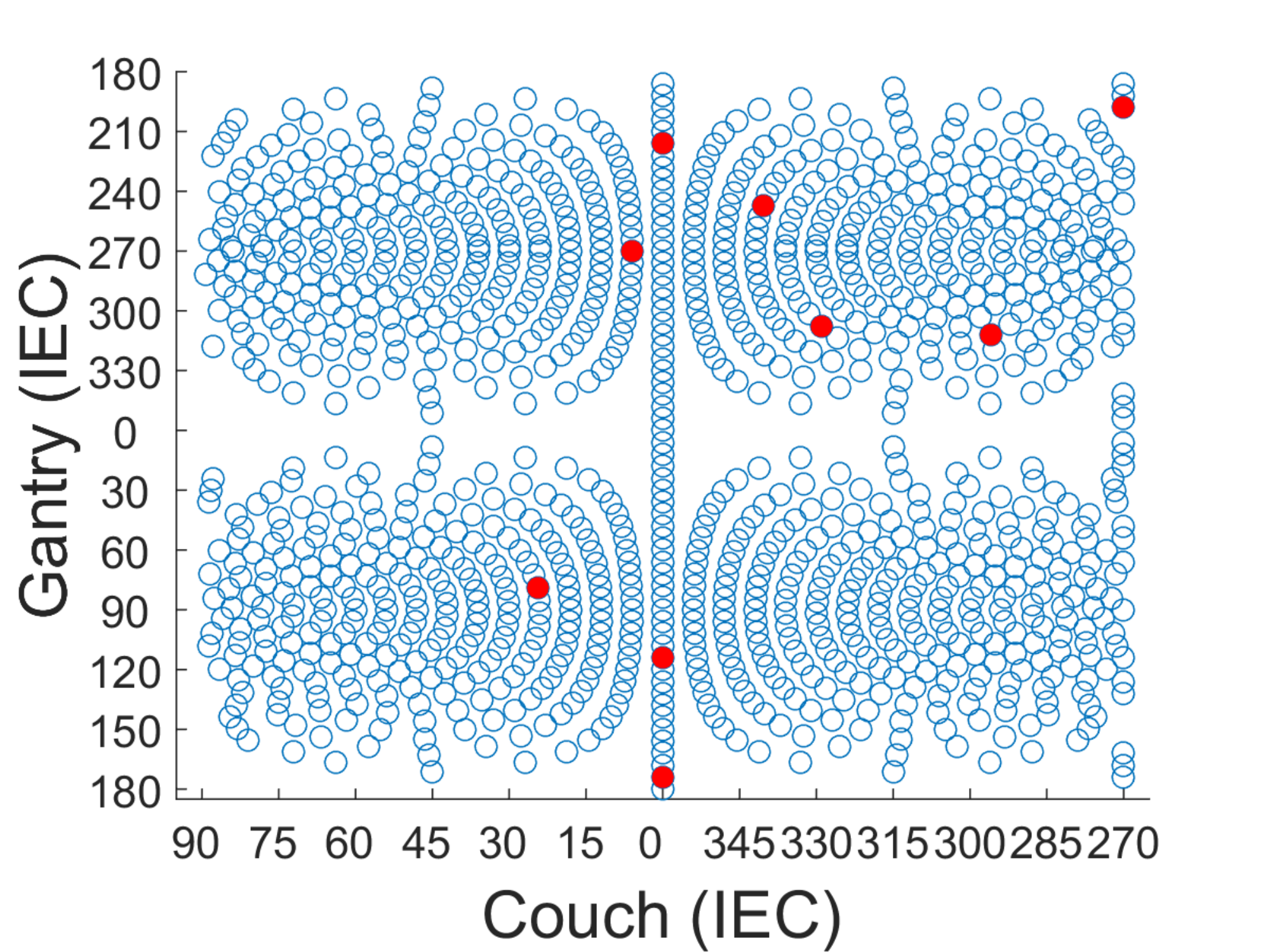}}\hfill
\subfloat[Fraction 4\label{prt_beams_frac4}]{\includegraphics[width=.3\textwidth]{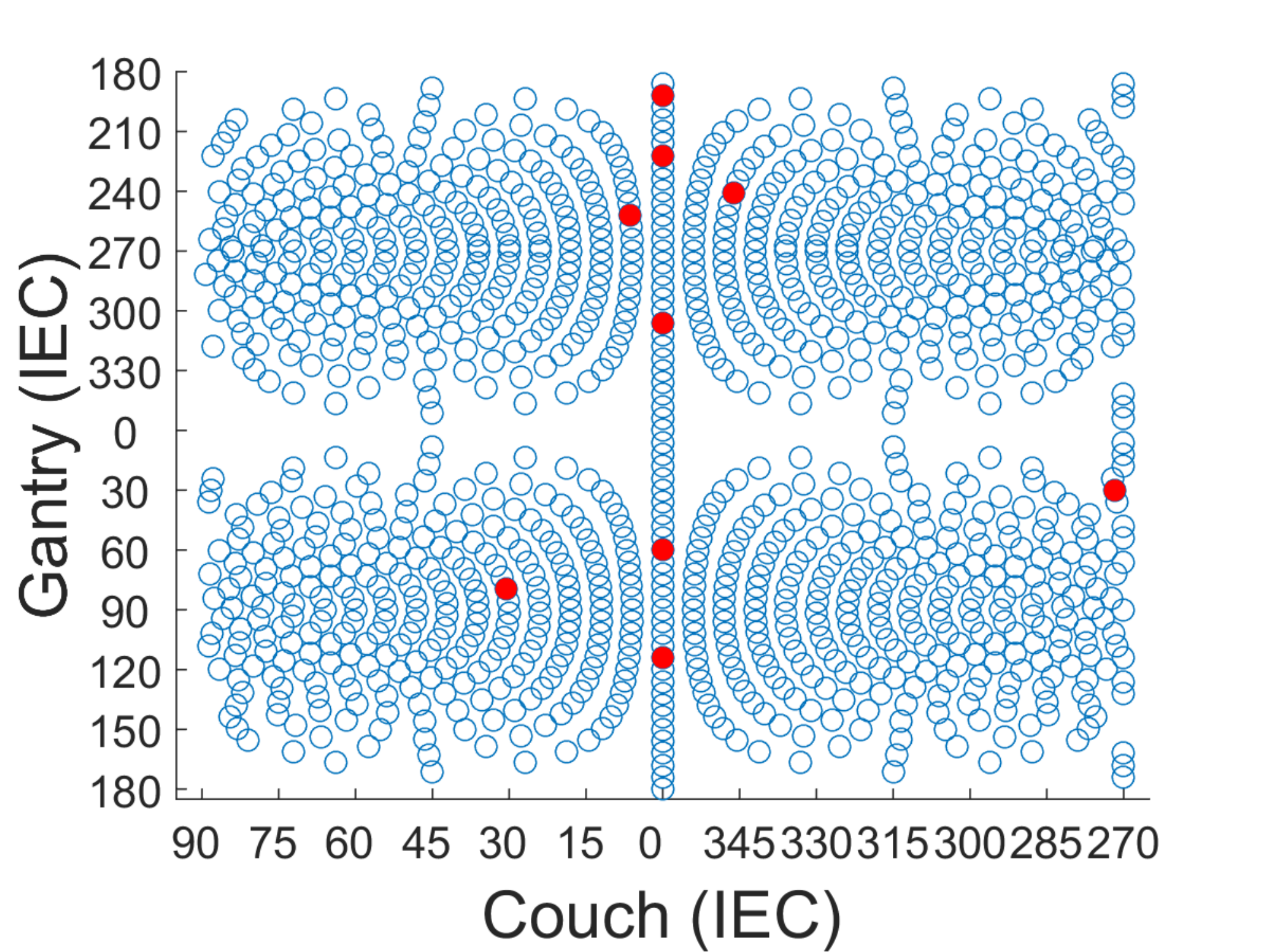}}\hfill
\subfloat[Fraction 5\label{prt_beams_frac5}]{\includegraphics[width=.3\textwidth]{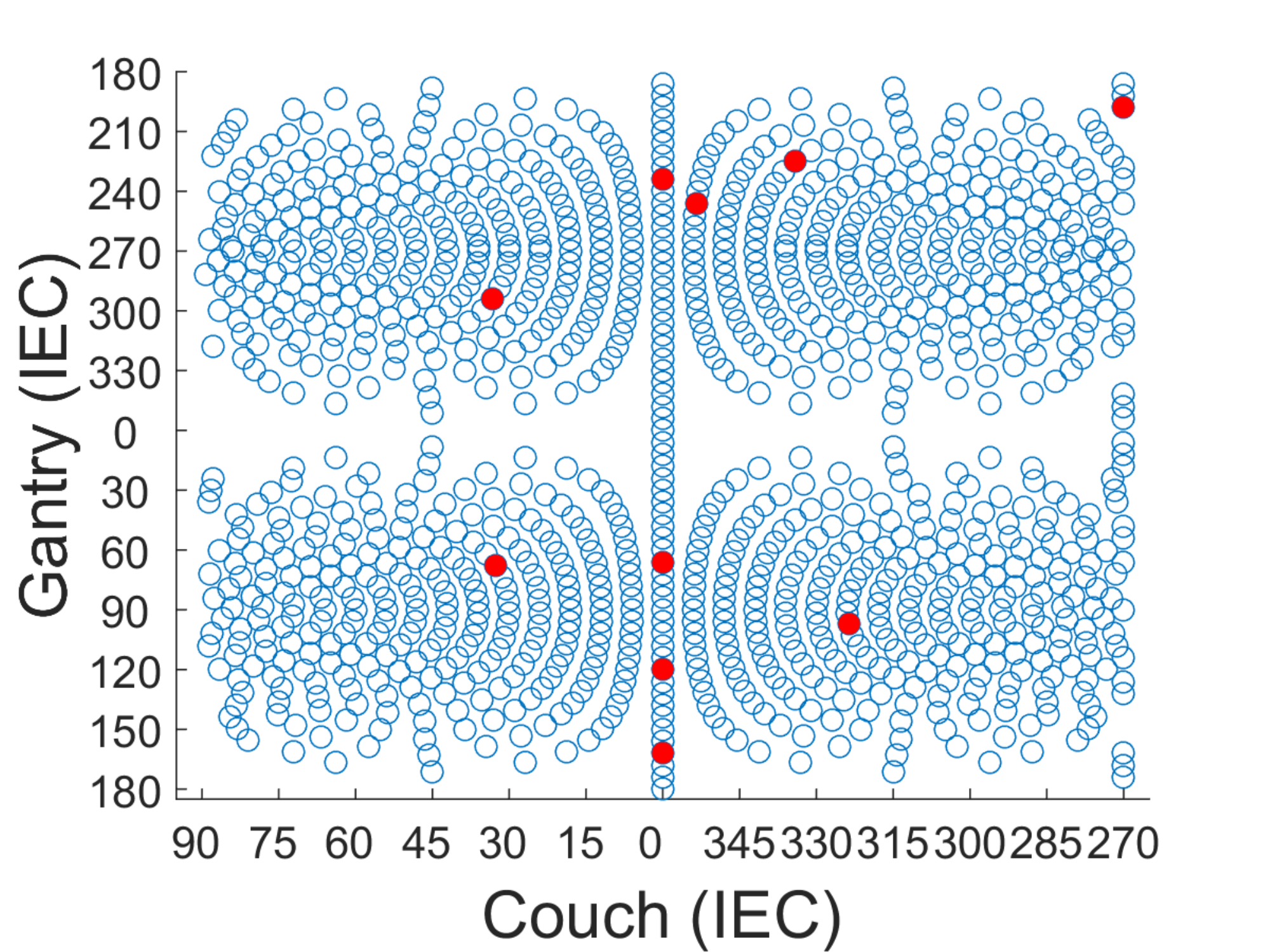}}\hfill
\subfloat[All fractions\label{prt_beams_allFracs}]{\includegraphics[width=.3\textwidth]{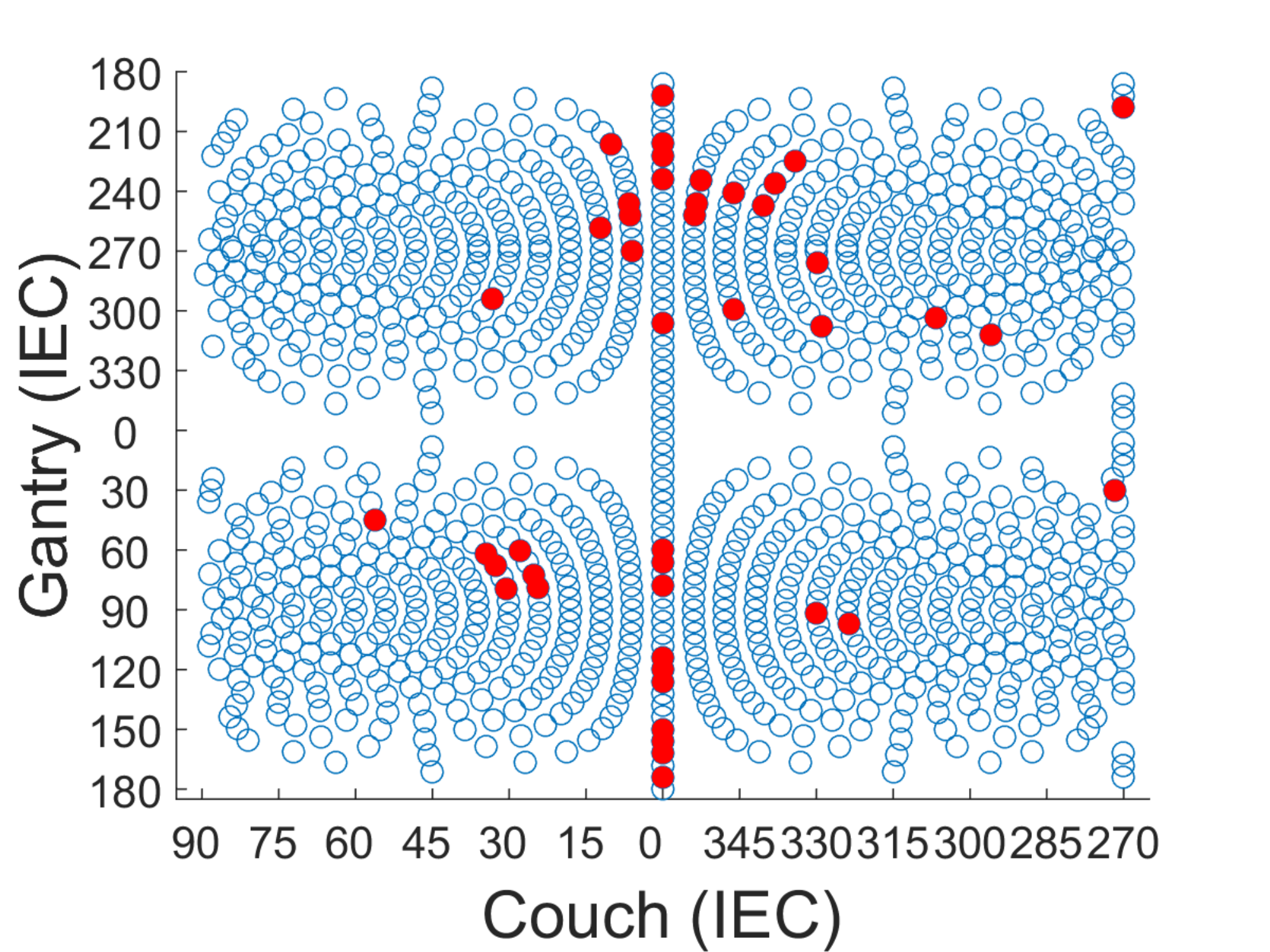}}
\caption{\protect\subref{prt_beams_frac1} - \protect\subref{prt_beams_frac5} Beam angles selected for each of five fractions for prostate case ``PRT''.
On average 9.6 beams per fraction were selected. \protect\subref{prt_beams_allFracs} A total of 44 distinct beam firing positions were utilized for case ``PRT''.}
\label{prt_beams}
\end{figure}

%%In the prostate case ``PRT'' there were 674 candidate beams per fraction,
%%for a total of 3370 candidate beams.
%%The total number of beamlets (including all candidate beams) was 569,770.
%%%After downsampling the voxel grid as described section~\ref{algorithmSection},
%%%the dose-calculation matrices $A_i$ took up a combined 6.9~GB of RAM.
%%The FISTA runtime to solve the FV BOO problem~\eqref{fracVariantProb} was 1.5 hours.
Figure~\ref{allFracsStacked}\subref{prt_allFracs} shows dose colormaps for each of the five fractions for the FV prostate plan.  
Notice that despite the variation in dose distributions, the PTV is covered uniformly in all fractions. 
Each voxel in the PTV receives a dose of approximately $40/5 = 8$ Gy per fraction. 
Figure~\ref{prt_beams} visualizes the beam angles that were selected for each of five fractions
for the prostate case using the FV BOO algorithm.
Each beam is specified by a couch angle and a gantry angle. 
%The fraction-variant
%BOO algorithm selected 10 beam angles for fractions~1 and~2, 9 beam angles
%for fractions~3 and~4, and 10 beam angles for fraction~5, for an average
%of 9.6 beams per fraction. 
The FV BOO algorithm selected 9.6 beams per fraction, on average.
As expected, the algorithm did not select
the same set of beam angles for each fraction.
In fact, a total of 44 distinct beam firing positions were utilized, as visualized
in figure~\ref{prt_beams}\subref{prt_beams_allFracs}.

%\begin{figure}[!htb]
%\centering
%\includegraphics[width=\textwidth]{D:/Daniel/fromAngelia/PRTAR/results/fractionated/nBeams_10_10_9_9_10/visBeams/bigImg_resize50percent_cloudConvert.eps}
%\caption{Beam angles selected for each of five fractions for prostate case ``PRT''.
%On average 9.6 beams per fraction were selected.}
%\label{PRTAR_visBeams}
%\end{figure}  %

%\begin{figure}[!htb]
%\centering
%\includegraphics[width=.5\textwidth]{D:/Daniel/fromAngelia/PRTAR/results/fractionated/nBeams_10_10_9_9_10/visBeams/all_44_beams.eps}
%\caption{A total of 44 distinct beam firing positions were utilized for case ``PRT''.}
%\label{PRTAR_allBeams}
%\end{figure}

% Notice the PROTECT command when referencing subfigures.  https://tex.stackexchange.com/a/62700/43391

\begin{figure}[!htb]
\centering
\includegraphics[width=.9\textwidth]{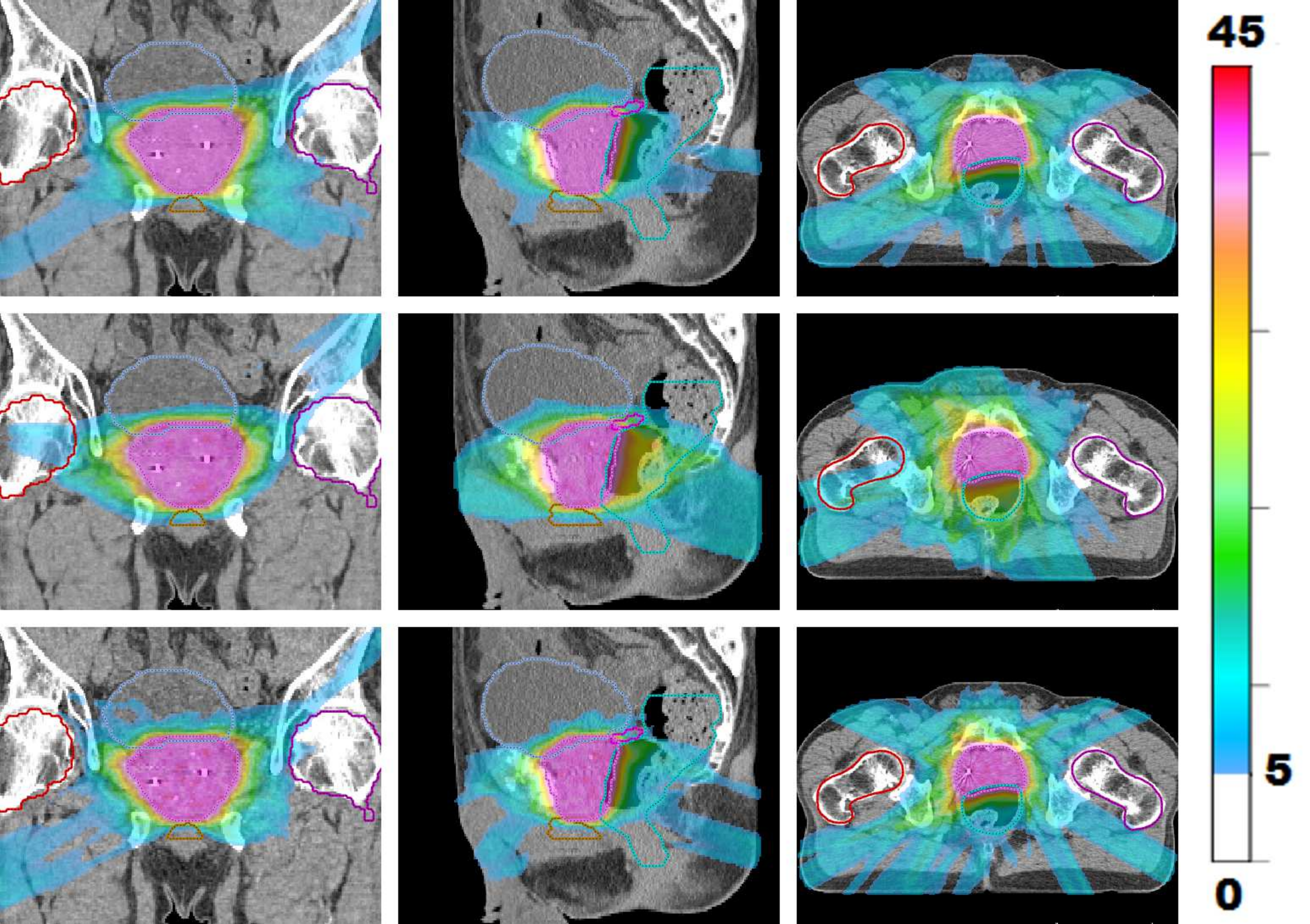}
\caption{Visualizing the dose distribution (summed over all five fractions) for FV plan using average of 9.6 beams per fraction (top row) as well as a 10-beam FI plan (middle row)
and a 20-beam FI plan (bottom row).
Dose below 5 Gy is not shown.}
\label{PRTAR_doseWash}
\end{figure}

\begin{figure}[!htb]
\centering
\subfloat[\label{prt_dvh_vs10beam}]{\includegraphics[width=.48\textwidth]{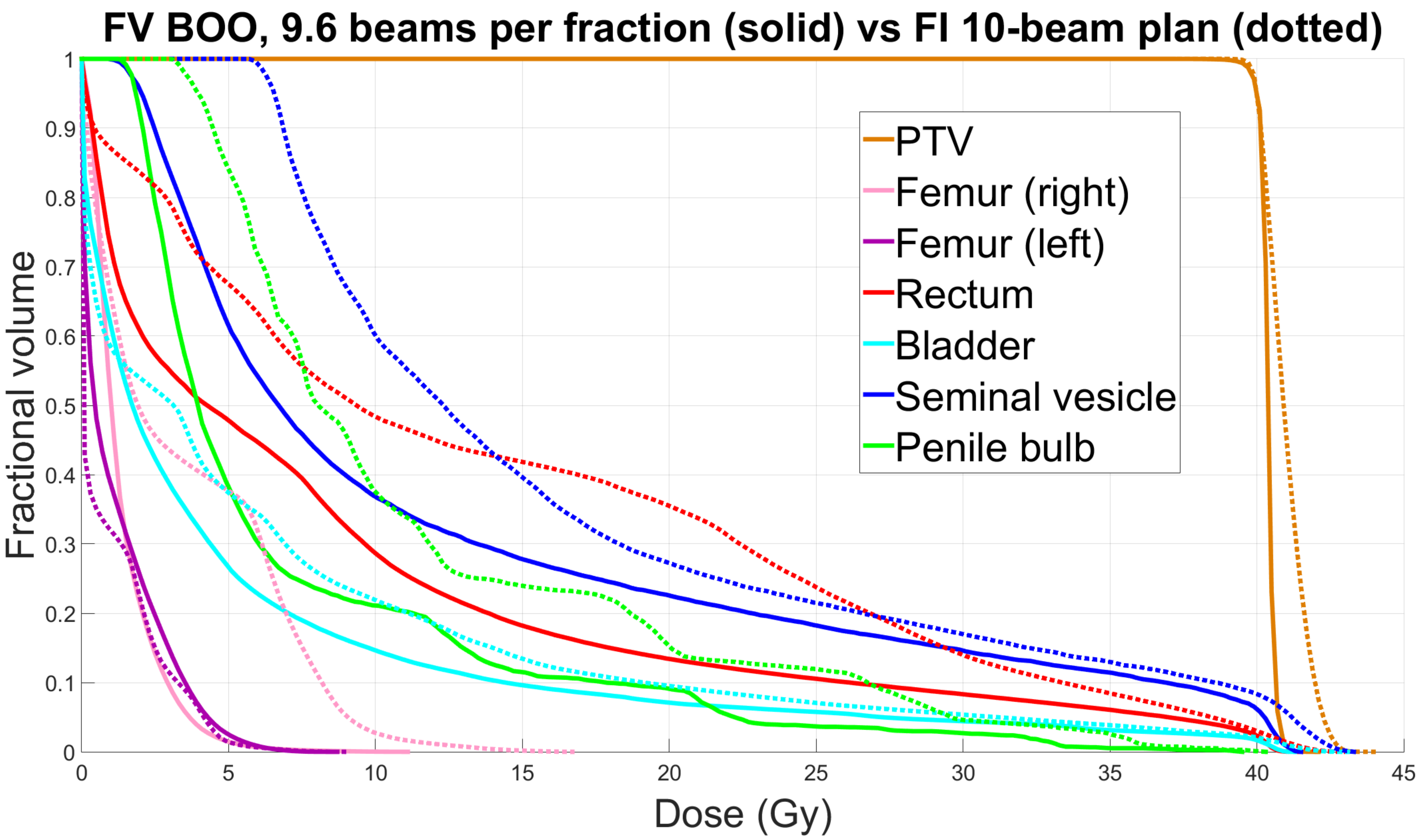}}\hfill
\subfloat[\label{prt_dvh_vs20beam}]{\includegraphics[width=.48\textwidth]{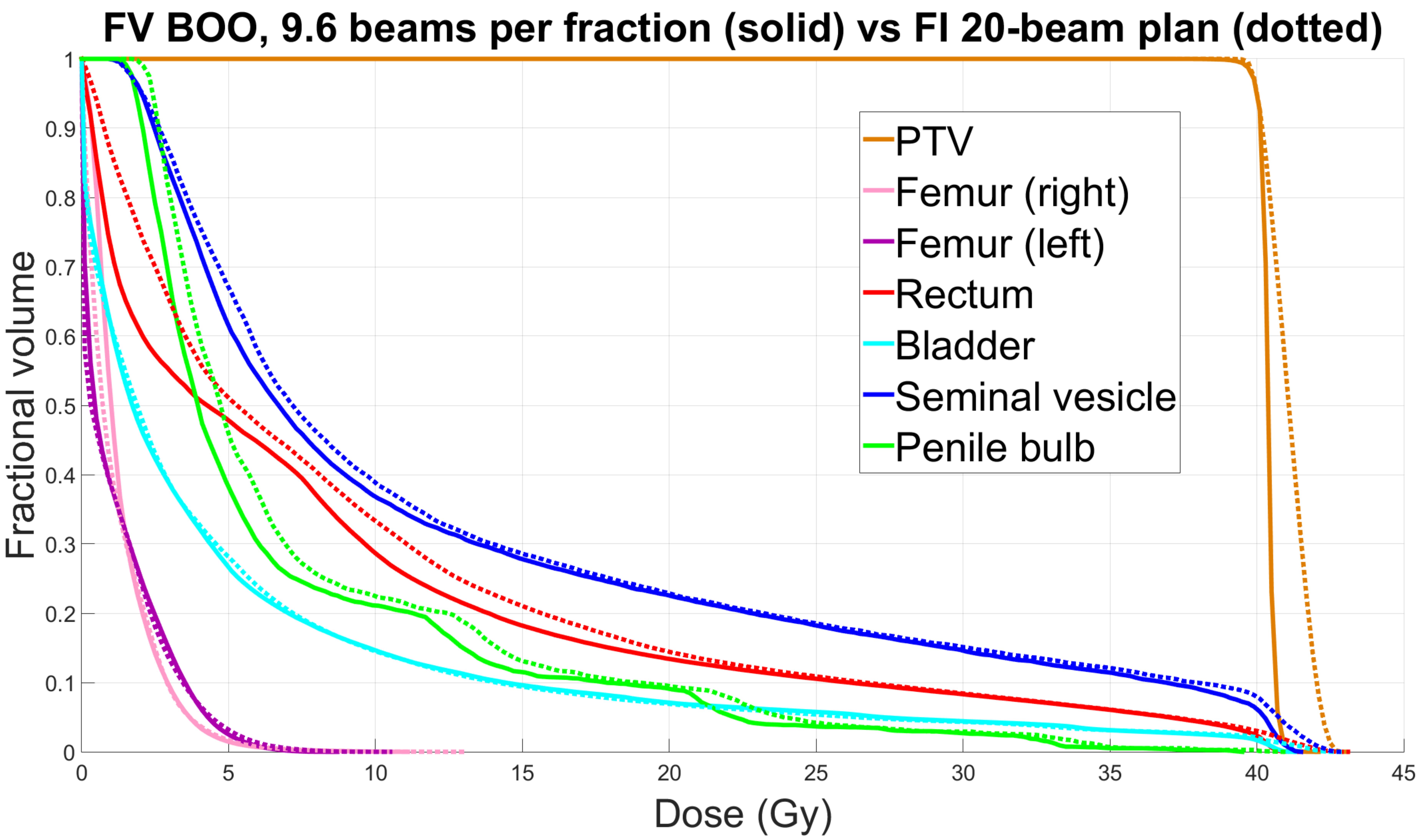}}
\caption{Dose volume histograms for prostate case ``PRT'', comparing FV plan that uses 9.6 beams per fraction (solid) with FI plans (dotted) using \protect\subref{prt_dvh_vs10beam} 10 beams,
and \protect\subref{prt_dvh_vs20beam} 20 beams.}
\label{prt_dvh}
\end{figure}

Figure~\ref{PRTAR_doseWash} shows the total dose distribution, summed over all five fractions,
for the FV plan (top row) as well as a conventional 10-beam FI plan
(middle row) and a conventional 20-beam FI plan (bottom row). 
Corresponding dose-volume histograms, comparing the FV plan
with the two FI beam plans, are shown in figure~\ref{prt_dvh}.
Compared with the 10-beam FI plan, the FV plan achieves dosimetric improvements
for all OARs except the left femur, which receives similar low doses in both plans.
The dosimetric improvement to the anterior rectum is particularly evident.
Mean dose was reduced by 5.7 Gy (rectum), 4.6 Gy (penile bulb),
4.3 Gy (seminal vesicle), and 1.4 Gy (bladder).
Max dose was reduced by 6.0 Gy (right femur), 3.3 Gy (penile bulb), and 1.3 Gy (seminal vesicle).
The PTV coverage is more homogeneous for the FV plan.
The dosimetric quality of the FV plan is similar to that of the 20-beam FI plan,
despite the fact that the FV plan uses only half as many beams per fraction.

%
%\begin{figure}[!htb]
%\centering
%\includegraphics[width=\textwidth]{D:/Daniel/fromAngelia/PRTAR/results/fractionated/nBeams_10_10_9_9_10/dvh_changed_fracBoo_vs_nonconvex10beams.eps}
%\caption{Dose volume histogram for prostate case ``PRT'', comparing FV plan that uses 9.6 beams per fraction (solid) with a 10-beam FI plan (dotted).}
%\label{PRTAR_dvh}
%\end{figure}

%\begin{figure}[!htb]
%\centering
%\includegraphics[width=\textwidth]{D:/Daniel/fromAngelia/PRTAR/results/fractionated/nBeams_10_10_9_9_10/dvh_changed_fracBoo_vs_nonconvex20beams.eps}
%\caption{Dose volume histogram for prostate case ``PRT'', comparing FV plan that uses 9.6 beams per fraction (solid)
%with a 20-beam FI plan (dotted).}
%\label{PRTAR_dvh_20beam}
%\end{figure}

%\begin{figure}[!htb]
%\centering
%\includegraphics[width=.9\textwidth]{D:/Daniel/fromAngelia/PRTAR/results/fractionated/nBeams_10_10_9_9_10/doseWashFracs/allFracs_withSpace.eps}
%\caption{Dose colormaps for each of five fractions for case ``PRT''. The PTV receives a uniform dose of $8$ Gy (approximately)
%at each fraction. Dose below .15 Gy is not shown.}
%\label{PRTAR_doseWash_allFracs}
%\end{figure}

\subsection{Lung case}

\begin{figure}[!htb]
\centering
\subfloat[Fraction 1\label{lng_beams_frac1}]{\includegraphics[width=.3\textwidth]{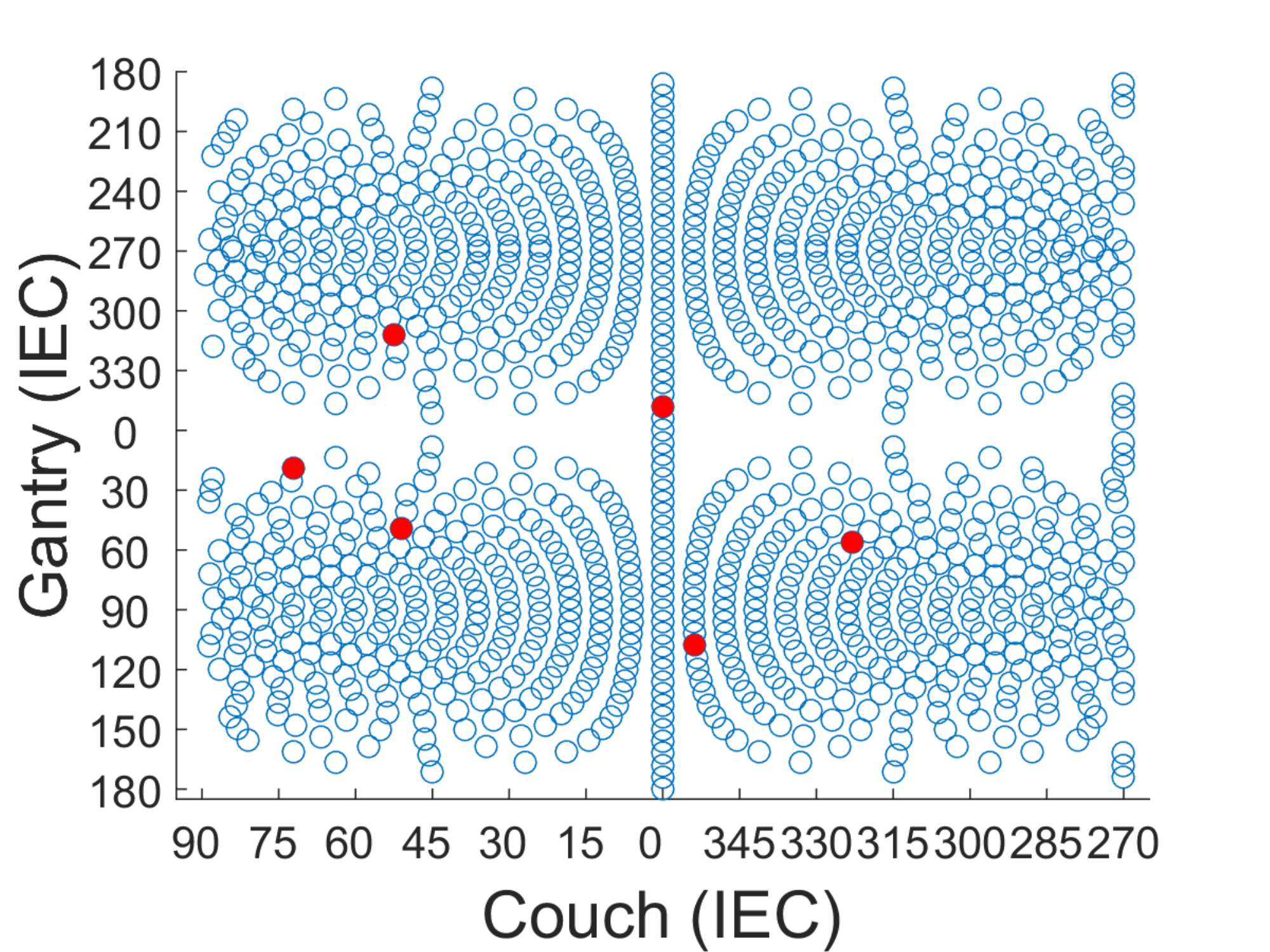}}\hfill
\subfloat[Fraction 2\label{lng_beams_frac2}]{\includegraphics[width=.3\textwidth]{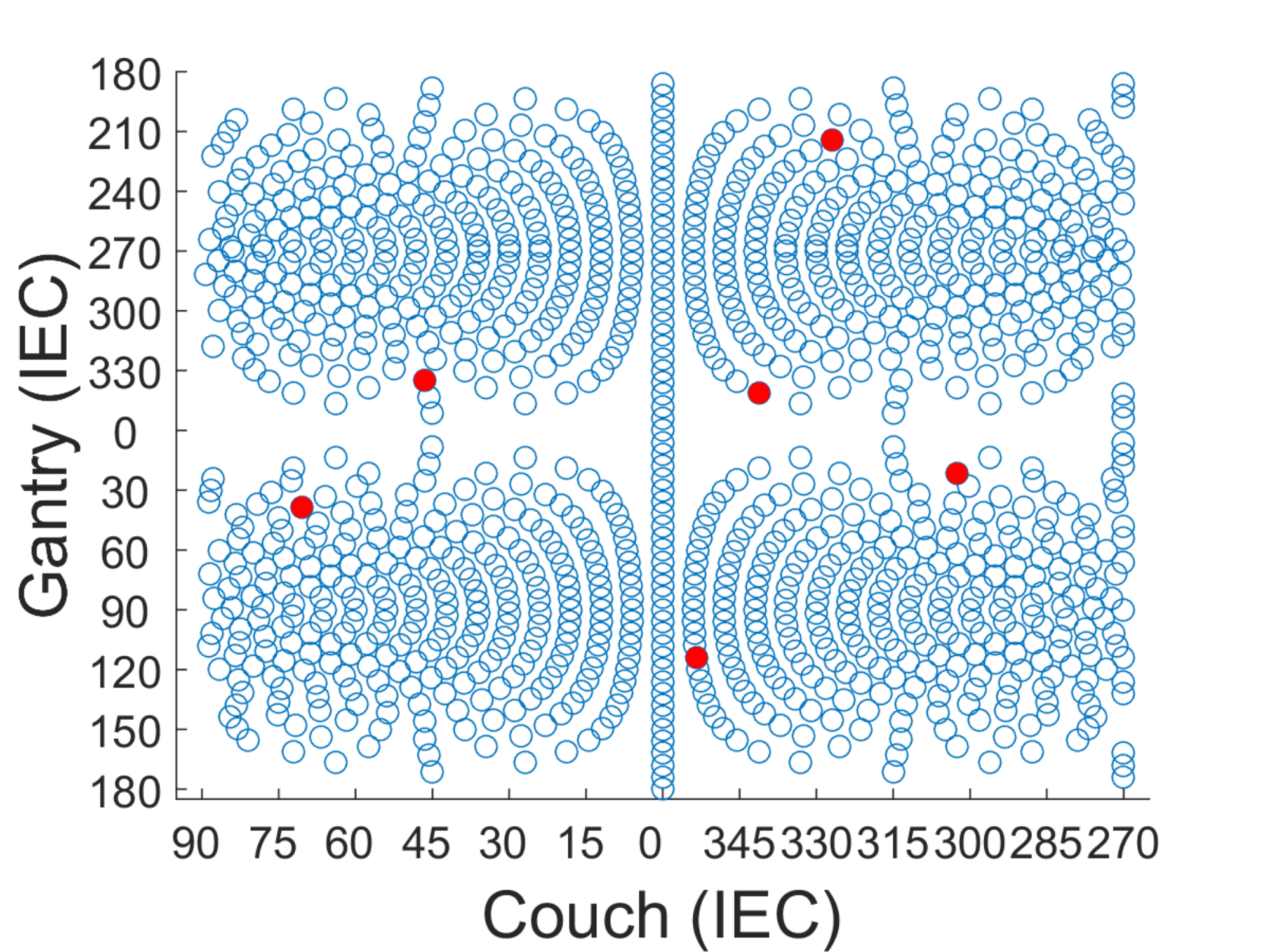}}\hfill
\subfloat[Fraction 3\label{lng_beams_frac3}]{\includegraphics[width=.3\textwidth]{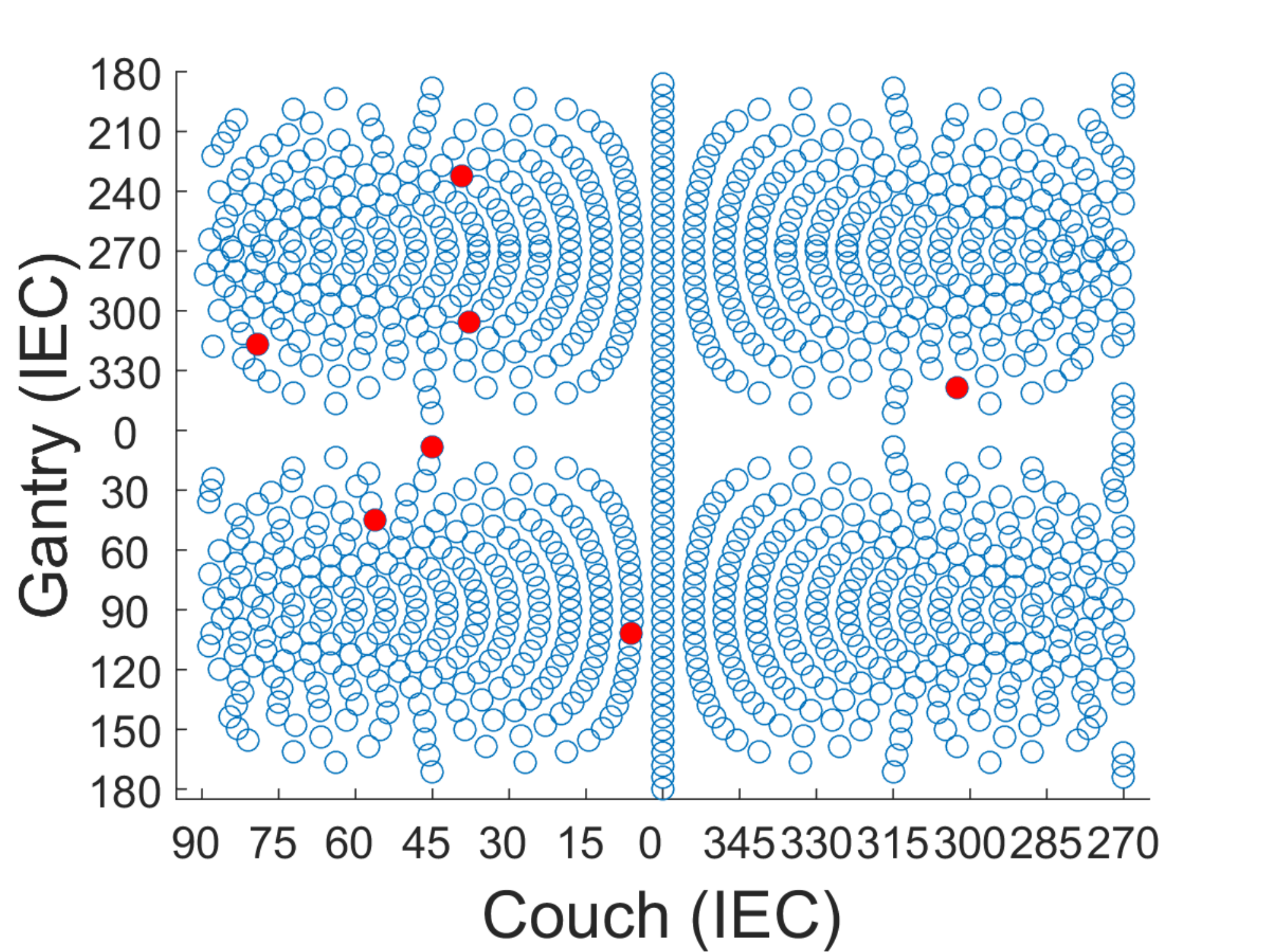}}\hfill
\subfloat[Fraction 4\label{lng_beams_frac4}]{\includegraphics[width=.3\textwidth]{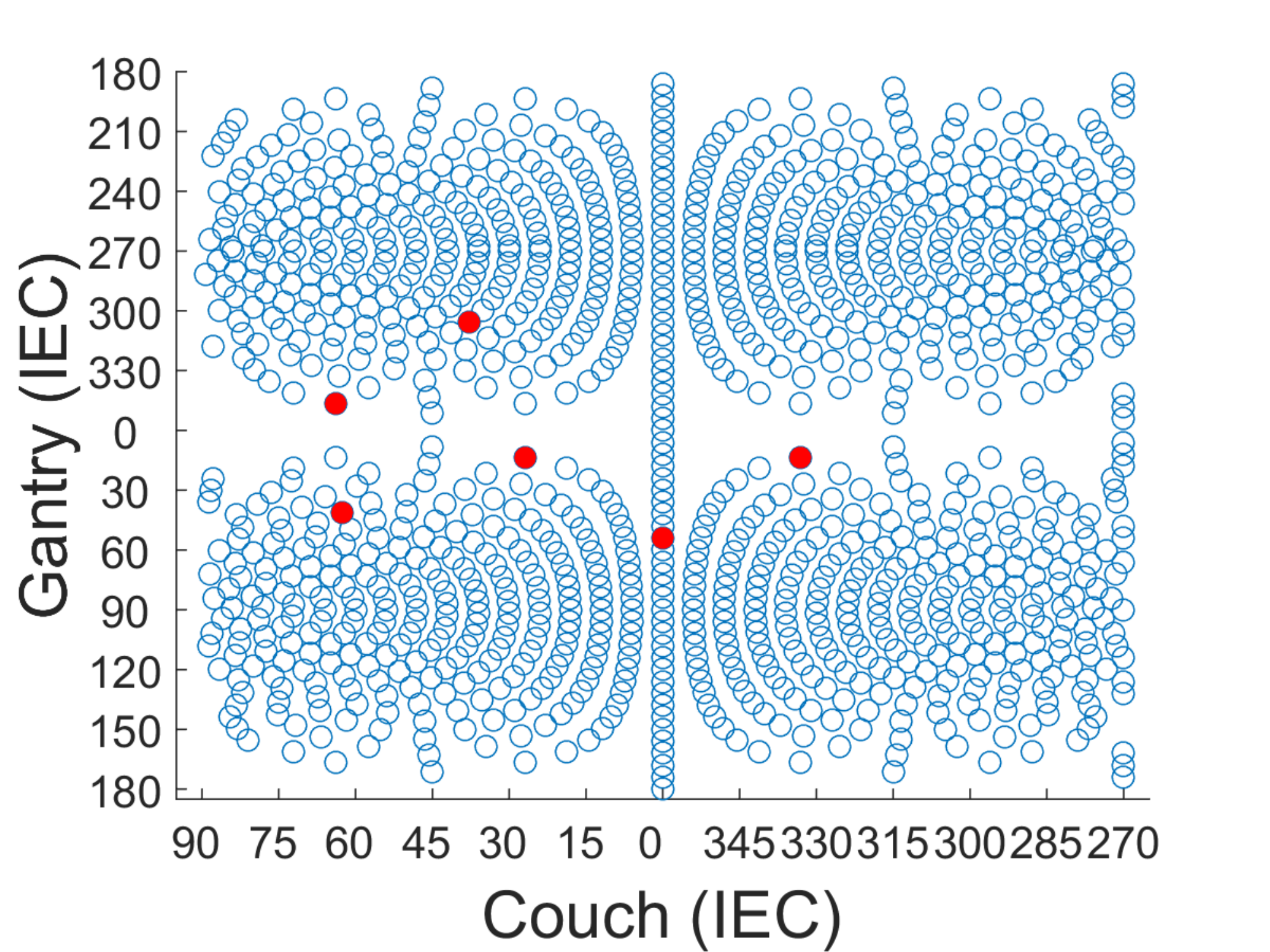}}\hfill
\subfloat[Fraction 5\label{lng_beams_frac5}]{\includegraphics[width=.3\textwidth]{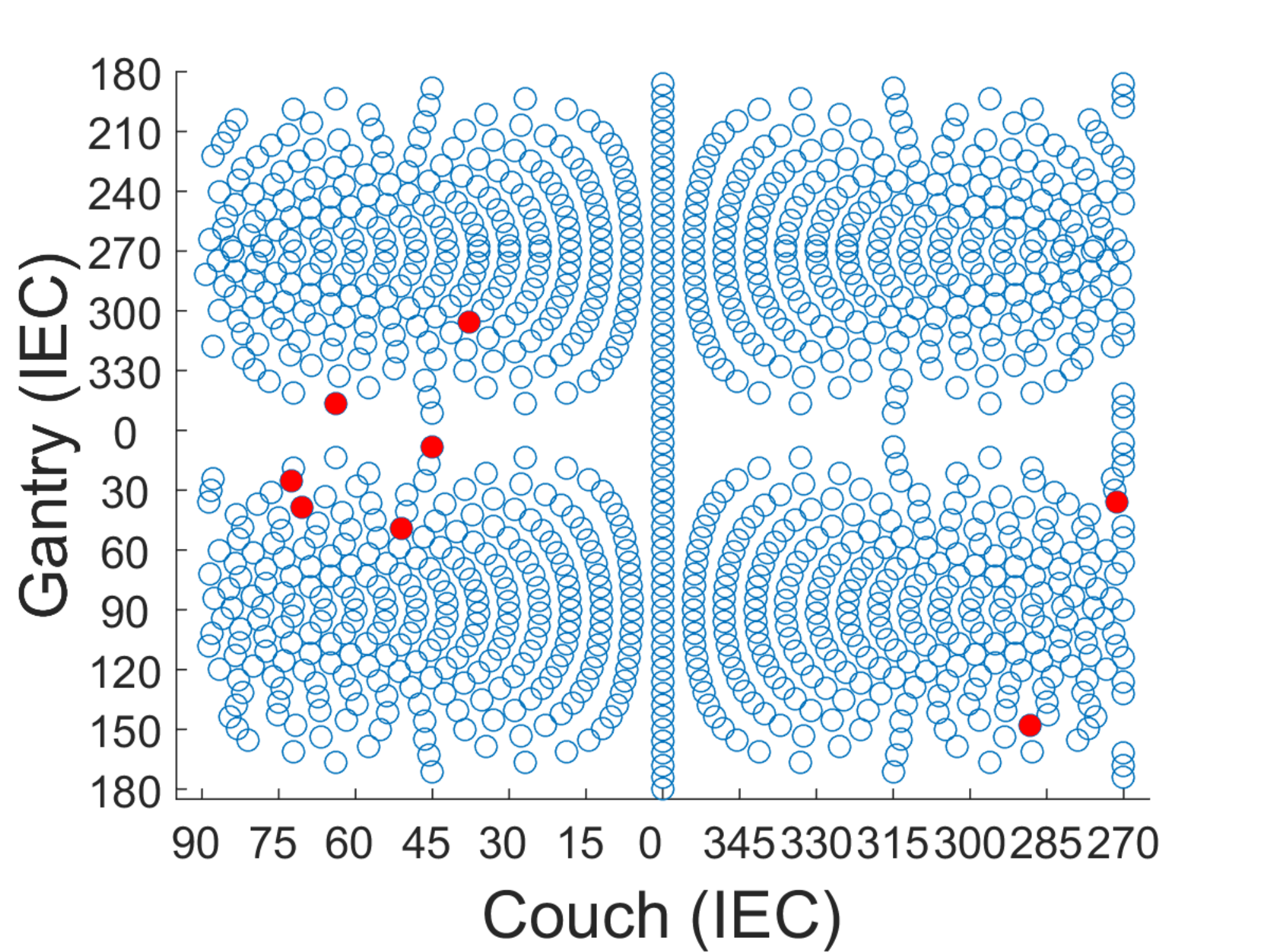}}\hfill
\subfloat[All fractions\label{lng_beams_allFracs}]{\includegraphics[width=.3\textwidth]{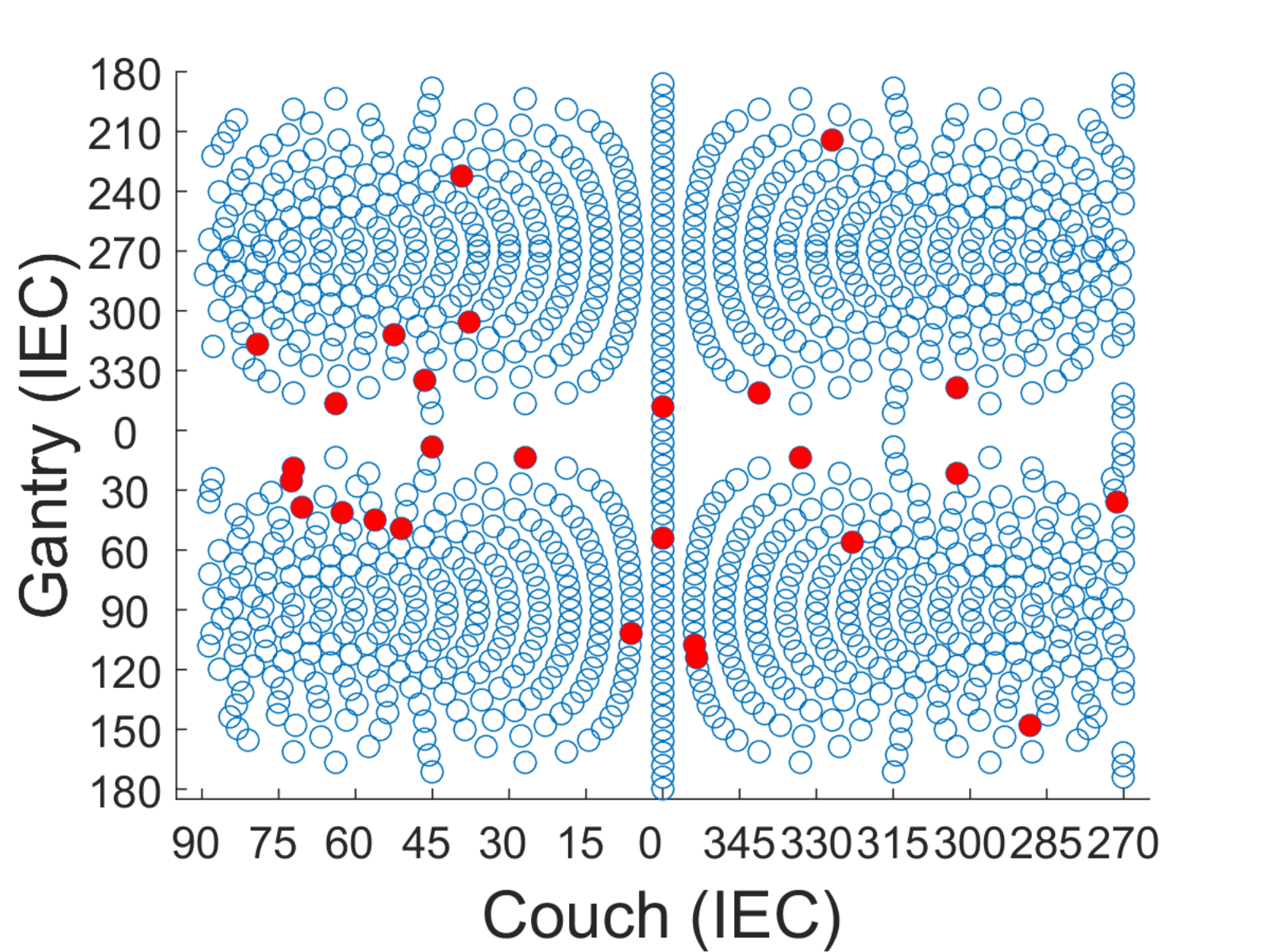}}
\caption{\protect\subref{lng_beams_frac1} - \protect\subref{lng_beams_frac5} Beam angles selected for each of five fractions for lung case ``LNG''.
On average 6.6 beams per fraction were selected. \protect\subref{lng_beams_allFracs} Although only 6.6 beams were selected per fraction (on average), a total of 27 distinct beam firing positions were utilized for case ``LNG''.}
\label{lng_beams}
\end{figure}

%%In the lung case ``LNG'' there were 520 candidate beams per fraction, for a total of 2600 candidate beams.
%%The total number of beamlets (including all candidate beams) was~$347,240$. 
%%%After downsampling the voxel grid, the dose-calculation matrices $A_i$ took up a combined $4.3$~GB of RAM.
%%The FISTA runtime to solve the FV BOO problem~\eqref{fracVariantProb} was 1.1 hours.
Figure~\ref{allFracsStacked}\subref{lng_allFracs} shows dose colormaps for each of the five fractions
for the FV lung plan.  
Despite the variation in dose distributions, the PTV is covered uniformly at each fraction. 
Each voxel in the PTV receives a dose of approximately $48/5 = 9.6$ Gy per fraction. 
Figure~\ref{lng_beams} shows the beam angles that were selected for each of five fractions for the lung case using the FV BOO algorithm.
On average, only 6.6 beams per fraction were selected. 
Again, the algorithm did not select the same set of beam angles for any two fractions.
A total of 27 distinct beam firing positions were utilized, as visualized
in figure~\ref{lng_beams}\subref{lng_beams_allFracs}.

%\begin{figure}[!htb]
%\centering
%\includegraphics[width=\textwidth]{D:/Daniel/LNGTM_dose_files/results/fractionated/nBeams_6_6_7_6_8/visBeams/bigImg.eps}
%\caption{Beam angles selected for each of five fractions for lung case ``LNG''.
%On average 6.6 beams per fraction were selected.}
%\label{LNGTM_visBeams}
%\end{figure}  %

%\begin{figure}[!htb]
%\centering
%\includegraphics[width=.5\textwidth]{D:/Daniel/LNGTM_dose_files/results/fractionated/nBeams_6_6_7_6_8/visBeams/all_27_beams.eps}
%\caption{Although only 6.6 beams were selected per fraction (on average), a total of 27 distinct beam firing positions were utilized for case ``LNG''.}
%\label{LNGTM_allBeams}
%\end{figure}

\begin{figure}[!htb]
\centering
\includegraphics[width=.9\textwidth]{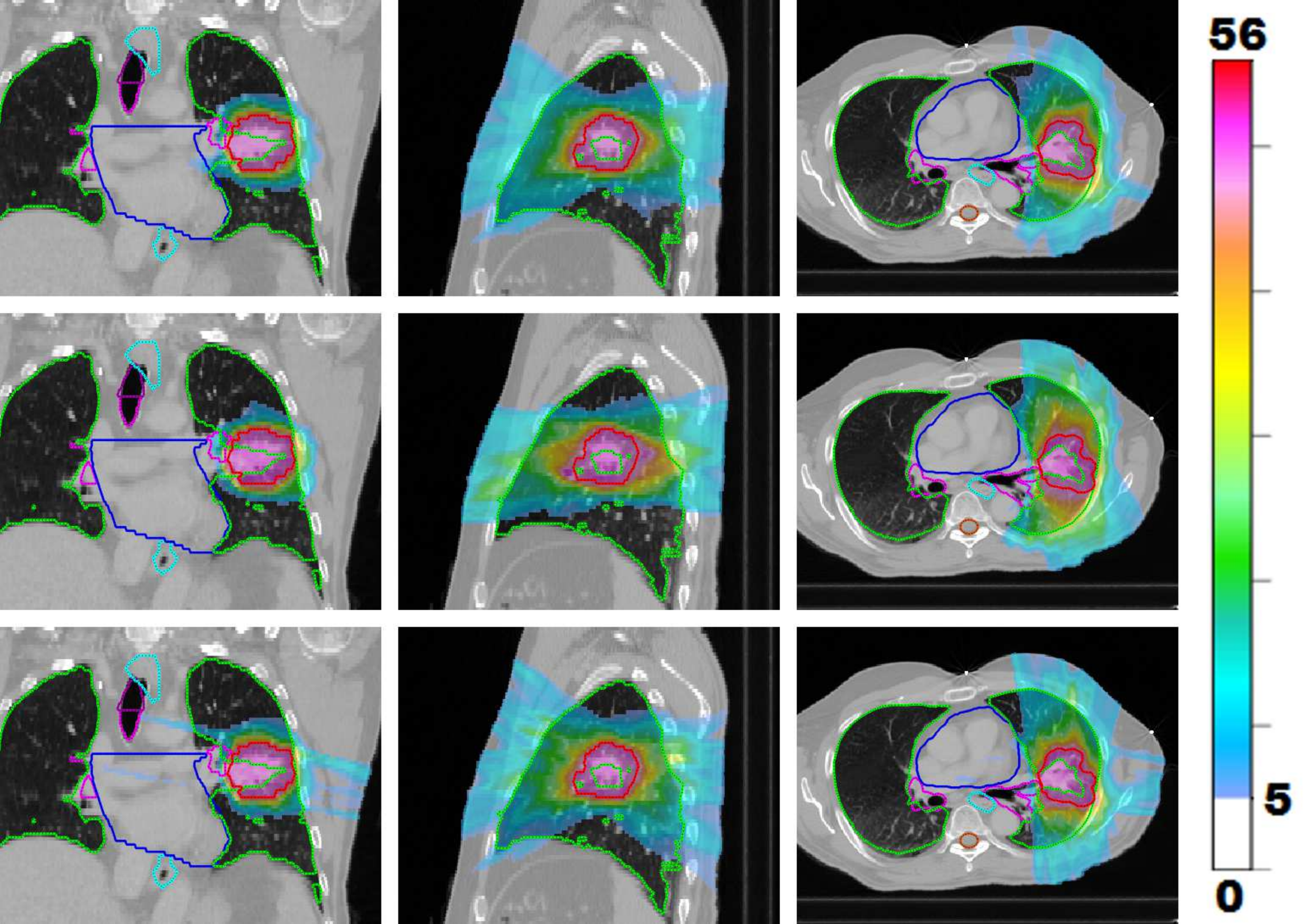}
\caption{The dose distribution (summed over all five fractions) for FV plan with 
average of 6.6 beams per fraction (top row) as well as a 7-beam FI plan (middle row)
and a 13-beam FI plan (bottom row).
Dose below 5 Gy is not shown.}
\label{LNGTM_doseWash}
\end{figure}

\begin{figure}[!htb]
\centering
\subfloat[\label{lng_dvh_vs7beam}]{\includegraphics[width=.48\textwidth]{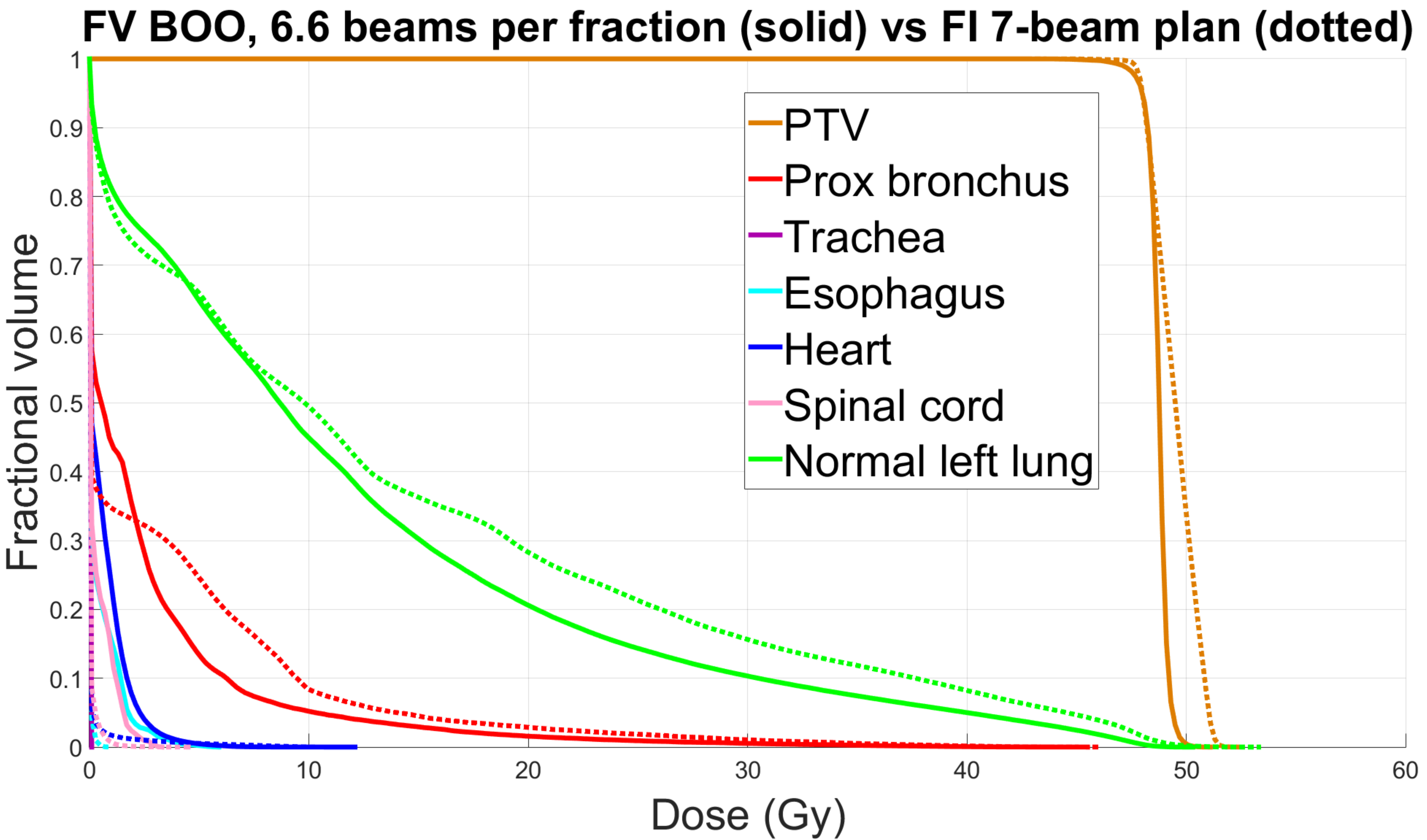}}\hfill
\subfloat[\label{lng_dvh_vs13beam}]{\includegraphics[width=.48\textwidth]{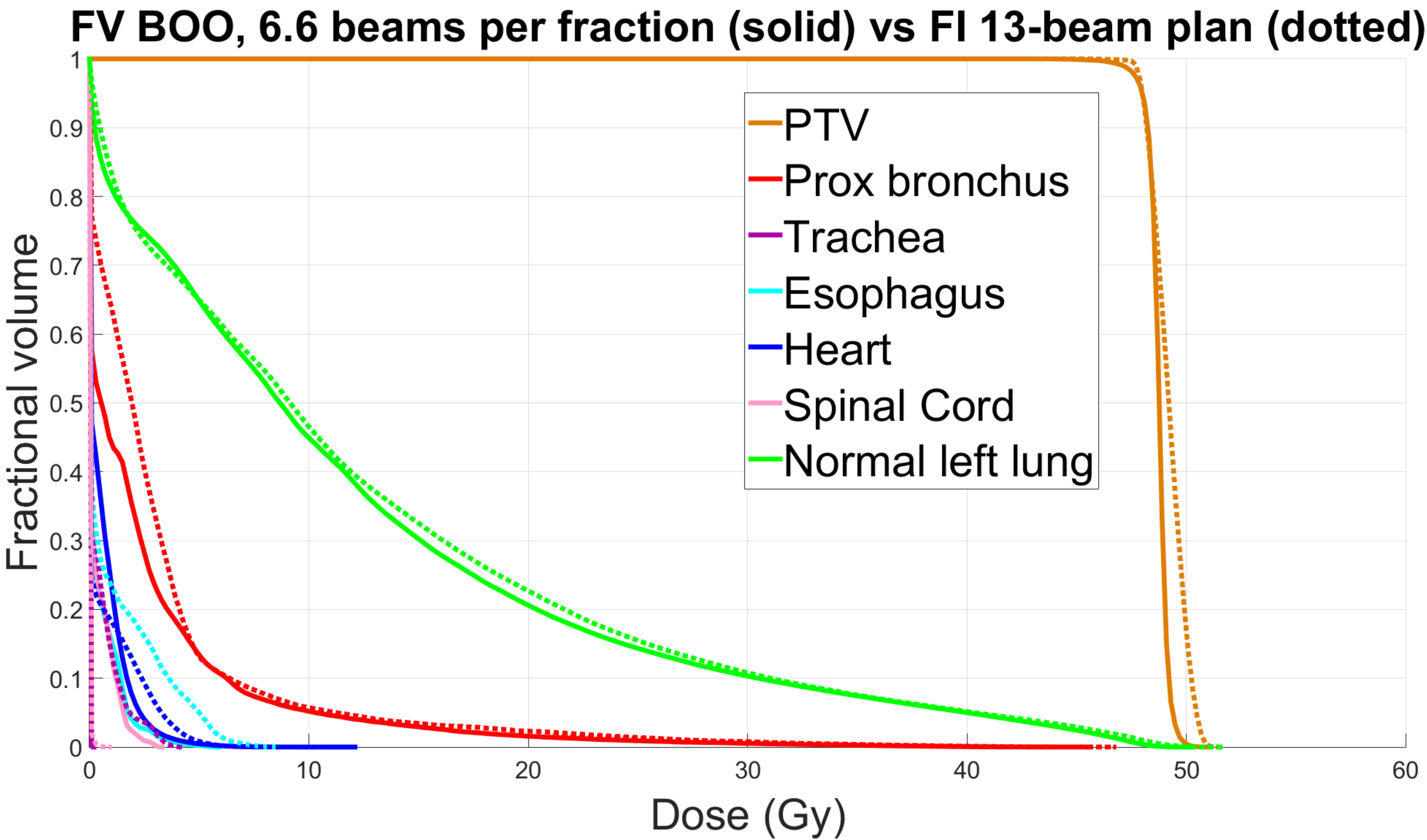}}
\caption{Dose volume histograms for lung case ``LNG'', comparing FV plan that uses 6.6 beams per fraction (solid) with FI plans (dotted) using \protect\subref{lng_dvh_vs7beam} 7 beams,
and \protect\subref{lng_dvh_vs13beam} 13 beams.}
\label{lng_dvh}
\end{figure}

Figure~\ref{LNGTM_doseWash} shows the total dose distribution, summed over all five fractions,
for the FV plan (top row) as well as a conventional 7-beam FI plan (middle row)
and a conventional 13-beam FI plan (bottom row).
Corresponding dose-volume histograms, comparing the FV plan
with the two FI plans, are shown in figure~\ref{lng_dvh}.
Compared with the 7-beam FI plan, the FV plan achieves dosimetric improvements
for the proximal bronchus and the normal left lung.
Mean dose was reduced by 1.9~Gy for the normal left lung,
while max dose was reduced by 6.0~Gy for the proximal bronchus
and by 2.0~Gy for the normal left lung.
The $R_{50}$ values are 2.98 for the FV plan, 4.12 for the 7-beam FI plan, and 3.25 for the 13-beam FI plan.
This substantial improvement in dose compactness for the FV plan can be visually appreciated in figure~\ref{LNGTM_doseWash}.
The dosimetric quality of the FV plan is similar to that of the 13-beam FI plan,
while using only half as many beams per fraction.

%\begin{figure}[!htb]
%\centering
%%\includegraphics[width=\textwidth]{D:/Daniel/LNGTM_dose_files/results/fractionated/nBeams_10_9_8_9_9/doseWashFracs/allFracs_betterColorbar.eps}
%\includegraphics[width=\textwidth]{D:/Daniel/LNGTM_dose_files/results/fractionated/nBeams_6_6_7_6_8/doseWashFracs/allFracs.eps}
%\caption{Dose colormaps for each of five fractions for case ``LNG''. The PTV receives a uniform dose of 9.6 Gy (approximately)
%at each fraction. Dose below .5 Gy is not shown.}
%\label{LNGTM_doseWash_allFracs}
%\end{figure}

\subsection{Head and neck case}  

\begin{figure}[!htb]
\centering
\subfloat[Fraction 1\label{hnk_beams_frac1}]{\includegraphics[width=.3\textwidth]{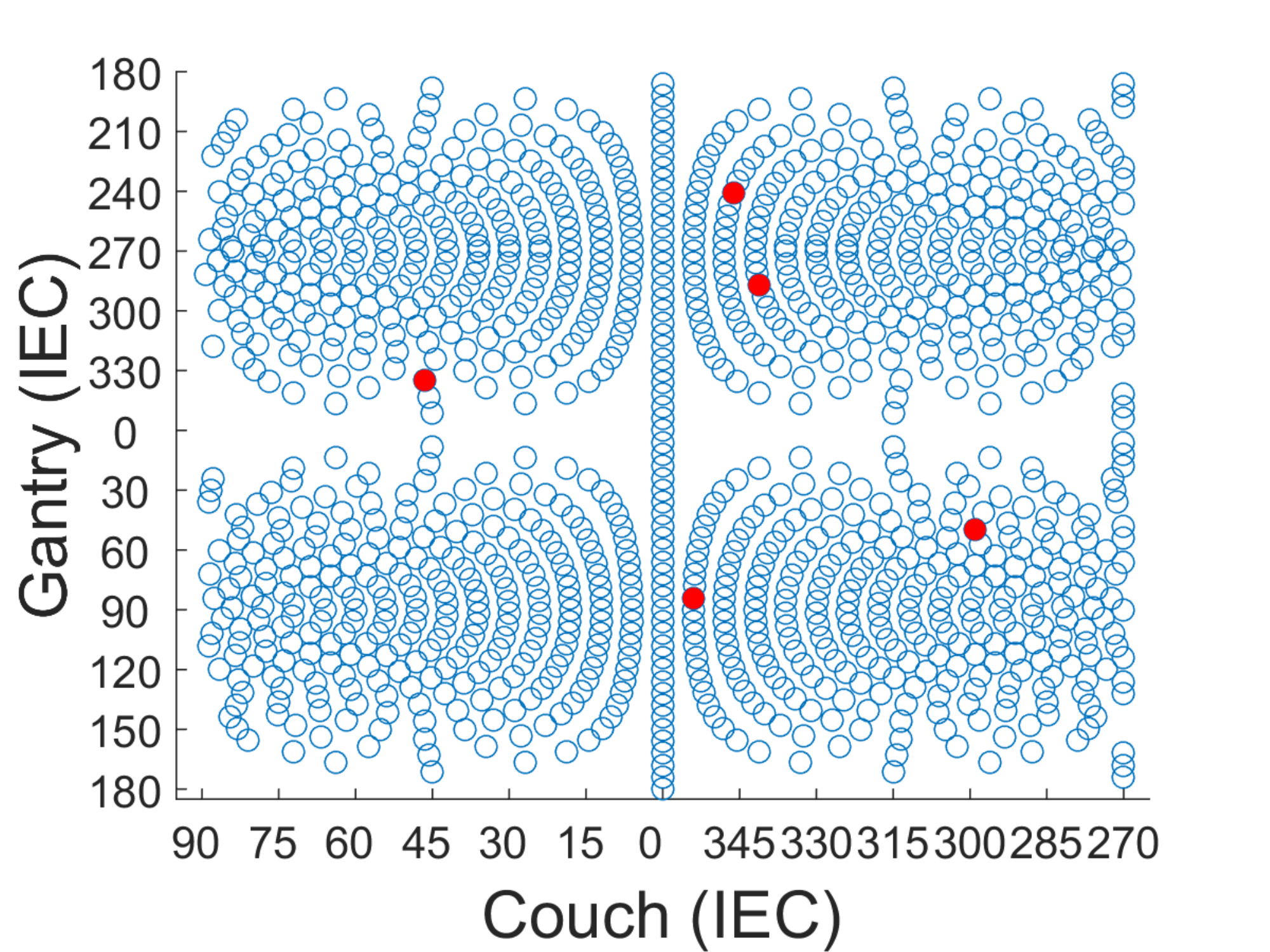}}\hfill
\subfloat[Fraction 7\label{hnk_beams_frac7}]{\includegraphics[width=.3\textwidth]{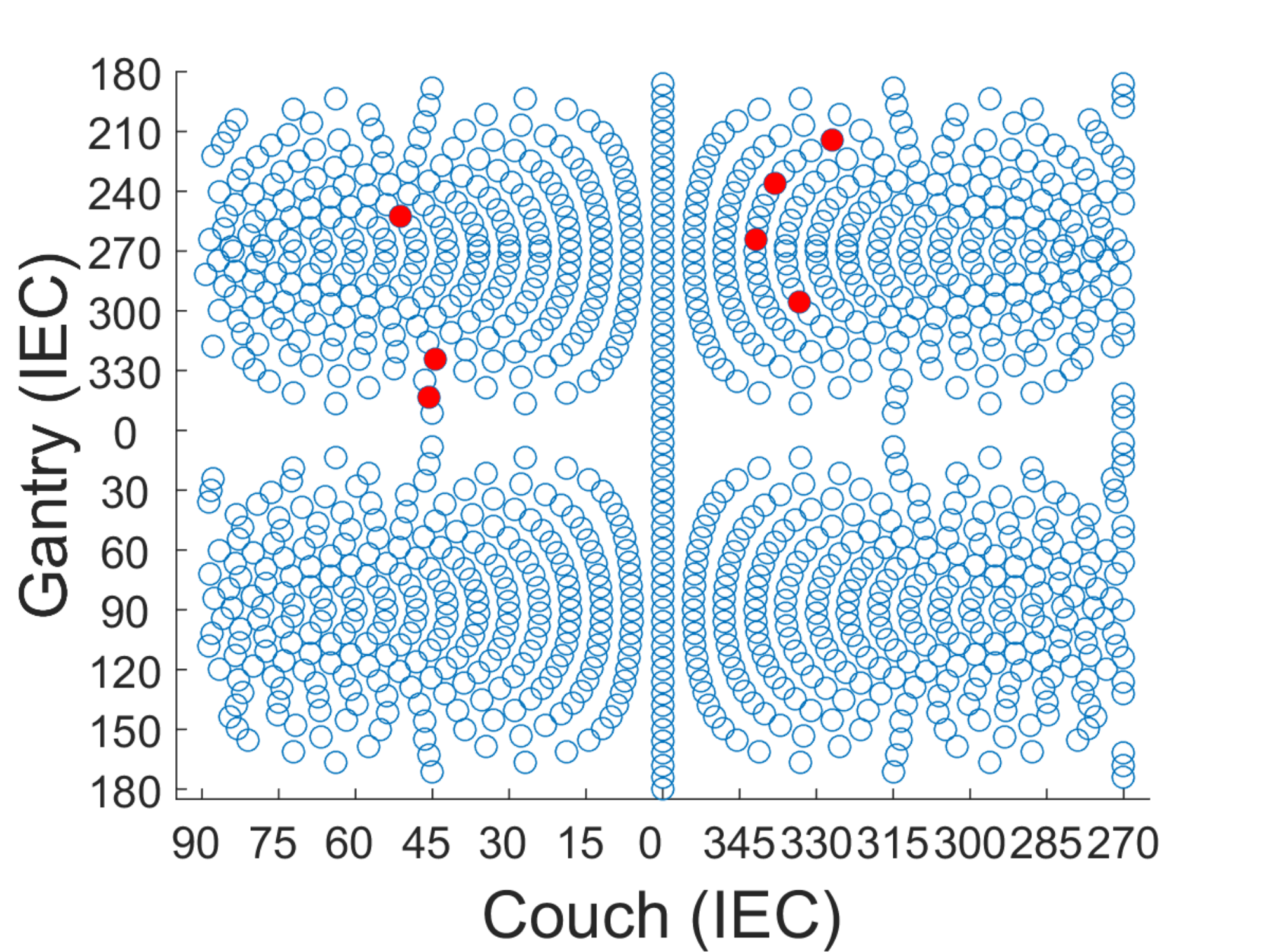}}\hfill
\subfloat[Fraction 13\label{hnk_beams_frac13}]{\includegraphics[width=.3\textwidth]{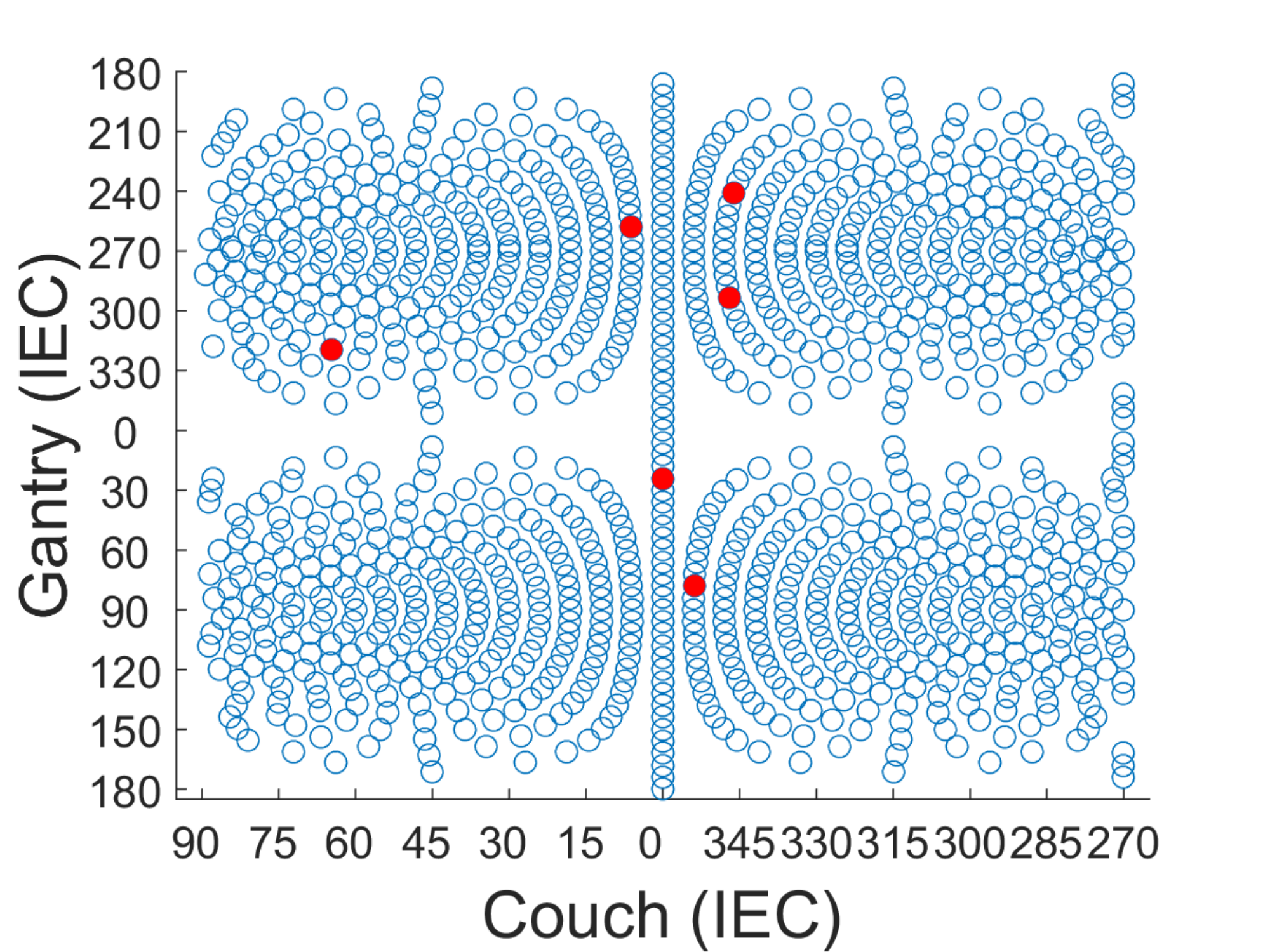}}\hfill
\subfloat[Fraction 19\label{hnk_beams_frac19}]{\includegraphics[width=.3\textwidth]{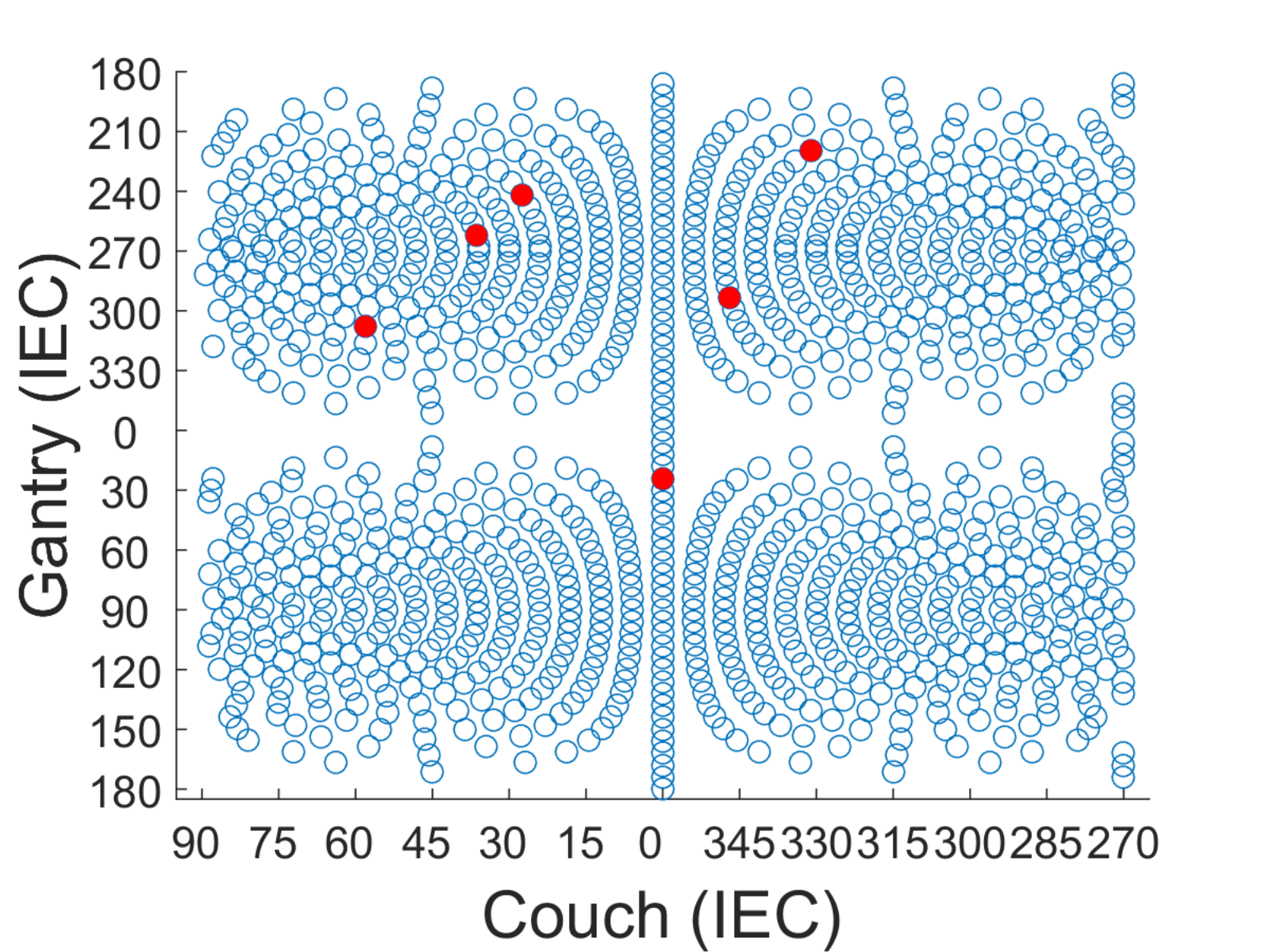}}\hfill
\subfloat[Fraction 25\label{hnk_beams_frac25}]{\includegraphics[width=.3\textwidth]{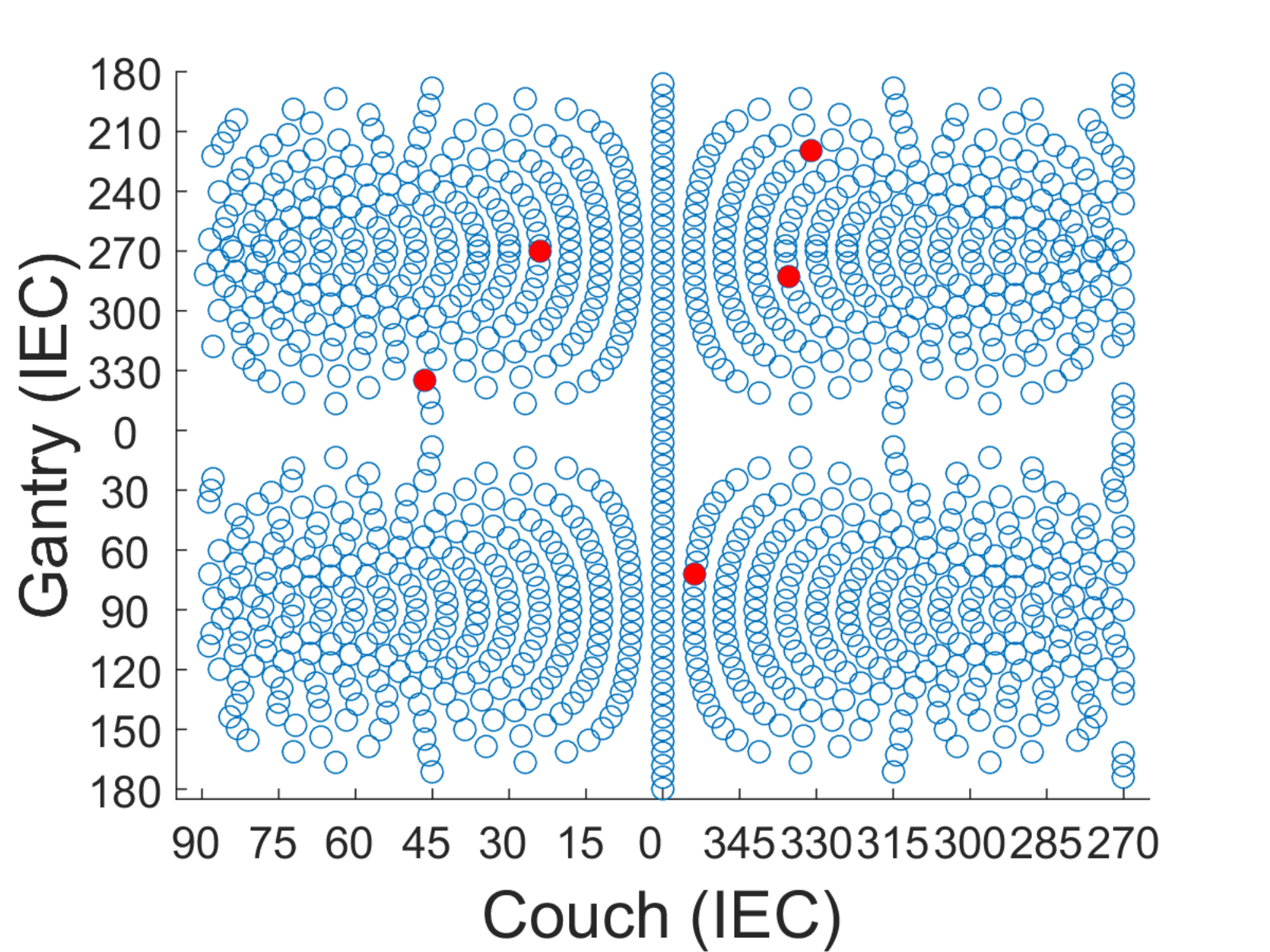}}\hfill
\subfloat[All fractions\label{hnk_beams_allFracs}]{\includegraphics[width=.3\textwidth]{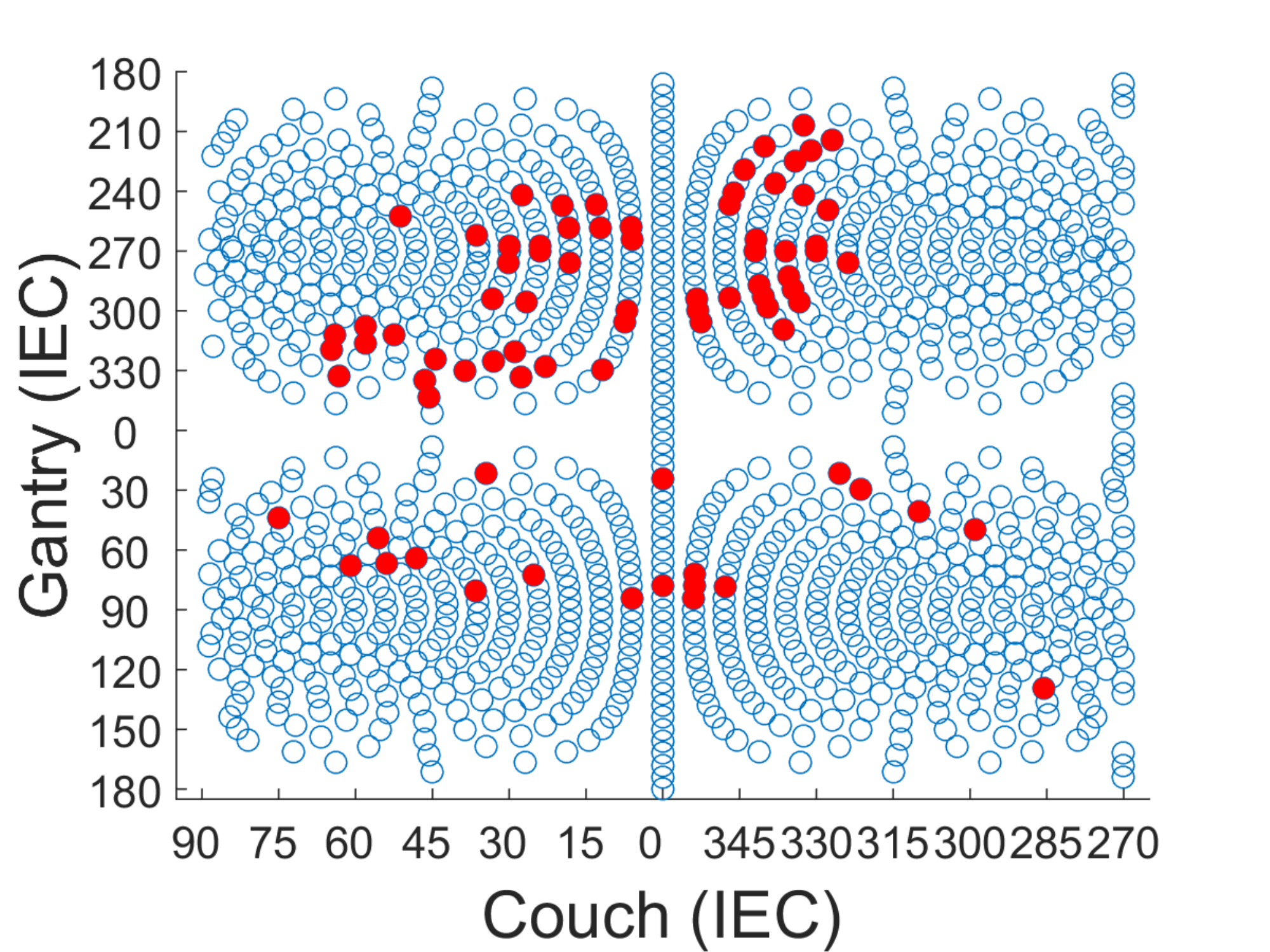}}
\caption{\protect\subref{hnk_beams_frac1} - \protect\subref{hnk_beams_frac25} Beam angles selected for five of the 30 treatment fractions for head and neck case ``H\&N''.
\protect\subref{hnk_beams_allFracs} Although only 6.36 beams were selected per fraction (on average), a total of 81 distinct beam firing positions were utilized for case ``H\&N''.}
\label{hnk_beams}
\end{figure}

%%In the head and neck case ``HNK'' there were 741 candidate beams per fraction, for a total of $741 \times 30 = 22,230$ candidate beams.
%%The total number of beamlets (including all candidate beams) was~2,013,870.
%%%After downsampling the voxel grid, the dose-calculation matrices $A_i$ took up a combined 9.0 GB of RAM.
%%The FISTA runtime to solve the fraction-variant BOO problem was 6.6 hours.
%Figure~\ref{HNKWL_visBeams} shows the beam angles that were selected for each of $30$ fractions
%for the head and neck case ``H\&N'' using the FV BOO algorithm.
Figure~\ref{allFracsStacked}\subref{hnk_allFracs} shows the dose distributions for a selection of five of the 30 treatment fractions
for the FV head and neck plan.  
%(Specifically, fractions 1, 7, 13, 19, and 25 are shown.)
Despite the variation in dose distributions,
the PTV is covered uniformly at each fraction. Each voxel in the PTV receives a dose of approximately $66/30 = 2.2$ Gy per fraction.
The FV BOO algorithm selected only $6.36$ beams per fraction, on average. 
As in all cases, the algorithm did not select the same set of beam angles for any two fractions.
A total of 81 distinct beam firing positions were utilized, as illustrated
in figure~\ref{hnk_beams}\subref{hnk_beams_allFracs}.

% XXX Reduce vertical space between top and bottom rows.

\begin{figure}[!htb]
\centering
\includegraphics[width=.9\textwidth]{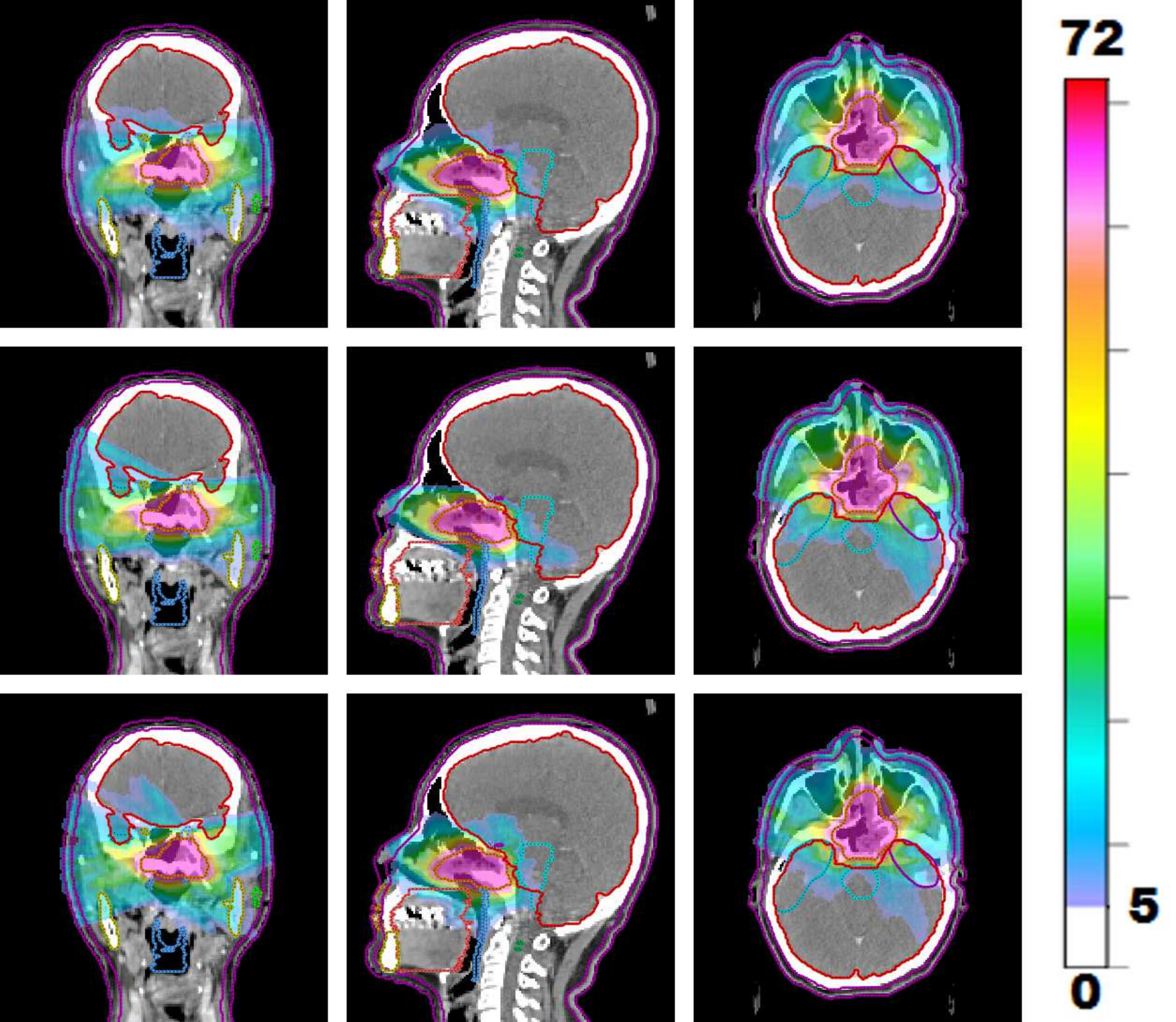}
\caption{Visualizing the dose distribution (summed over all 30 fractions) for FV plan with average of 6.36 beams per fraction (top row) as well as a 7-beam FI plan (middle row)
and a 13-beam FI plan (bottom row).
Dose below 5 Gy is not shown.}
\label{HNKWL_doseWash}
\end{figure}

\begin{figure}[!htb]
\centering
\subfloat[\label{hnk_dvh_vs7beam}]{\includegraphics[width=.48\textwidth]{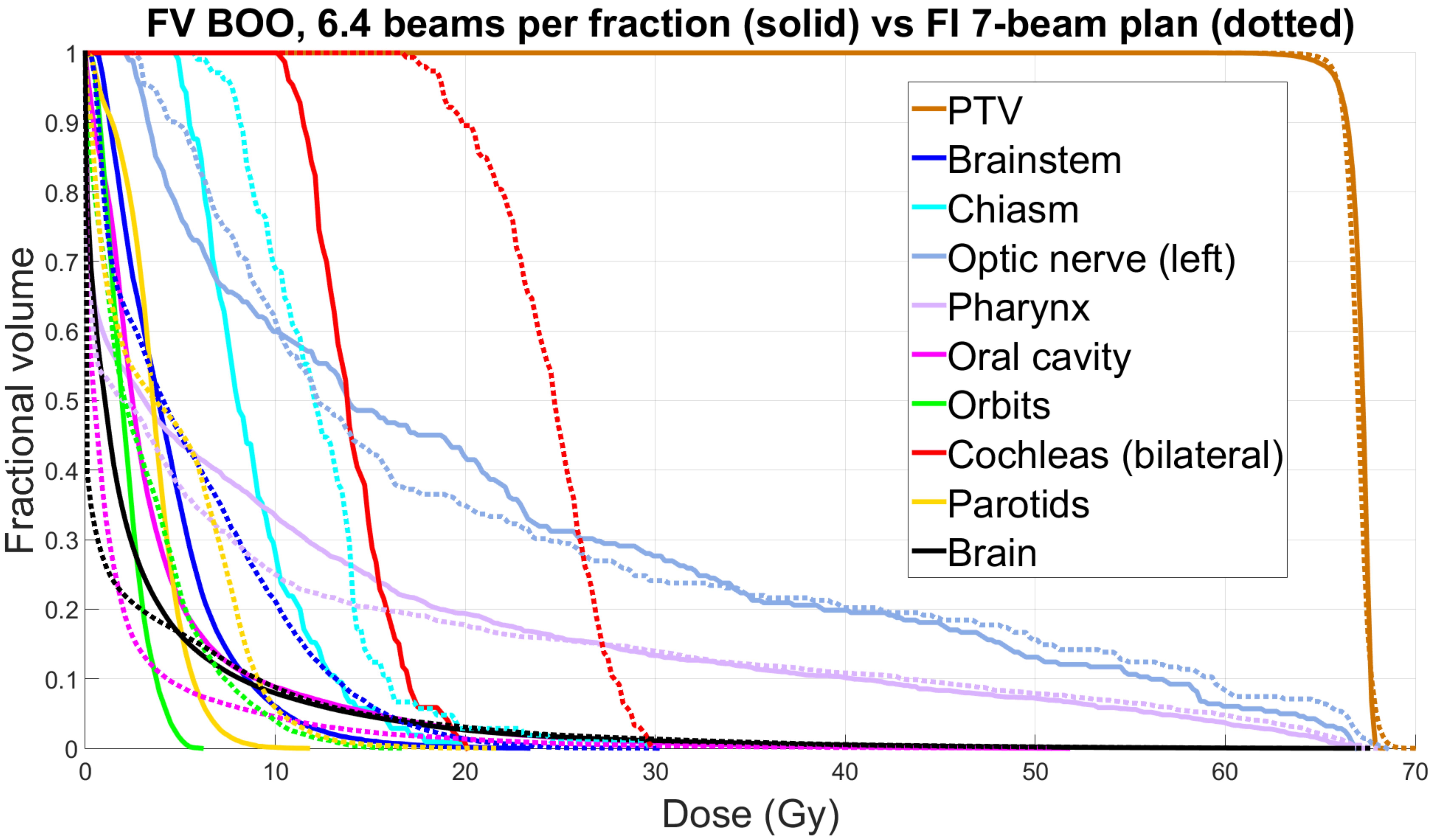}}\hfill
\subfloat[\label{hnk_dvh_vs13beam}]{\includegraphics[width=.48\textwidth]{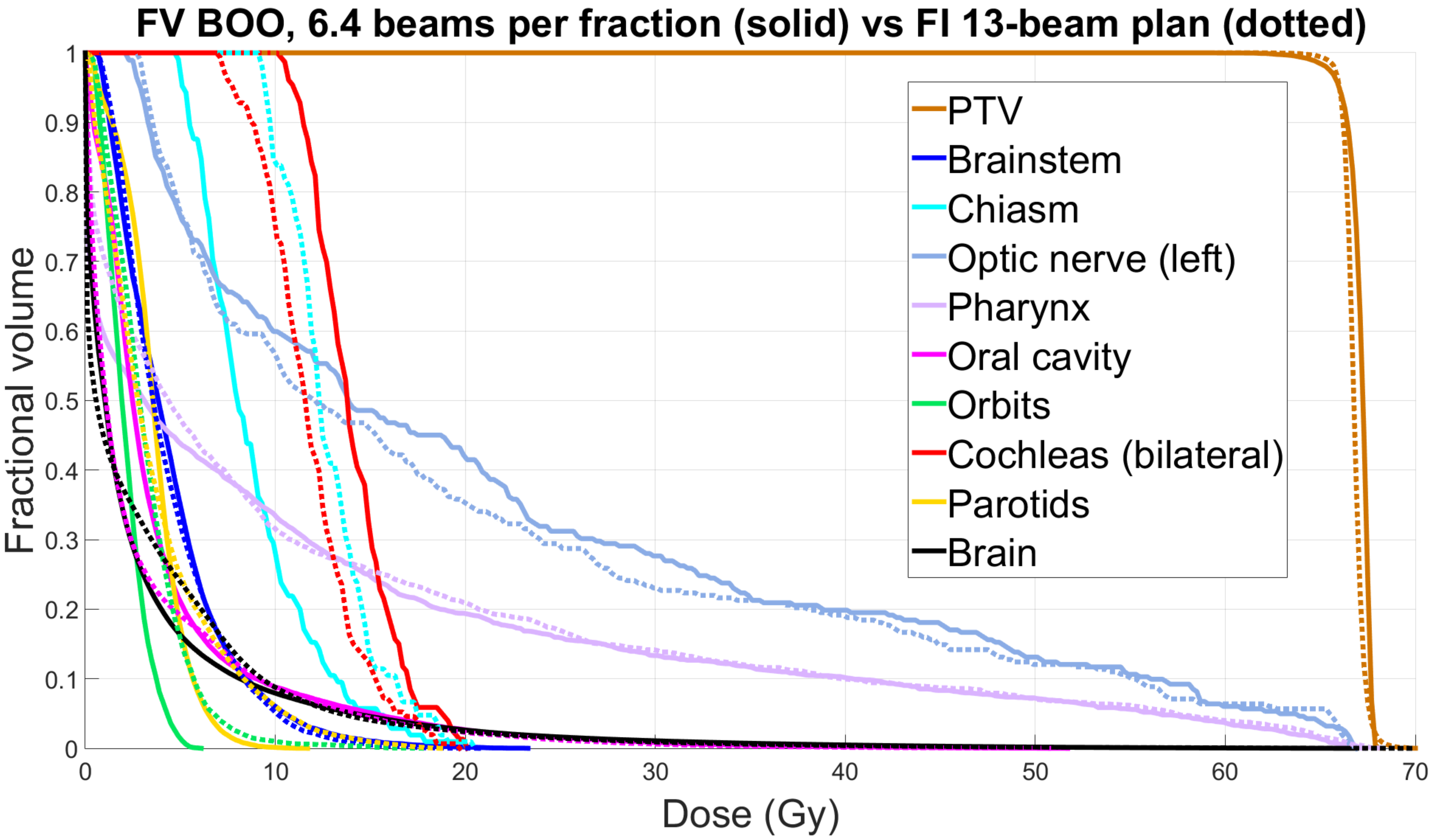}}
\caption{Dose volume histograms for head and neck case ``H\&N'', comparing FV plan that uses 6.36 beams per fraction (solid) with FI plans (dotted) using \protect\subref{hnk_dvh_vs7beam} 7 beams,
and \protect\subref{hnk_dvh_vs13beam} 13 beams.}
\label{hnk_dvh}
\end{figure}

Figure~\ref{HNKWL_doseWash} shows the total dose distribution, summed over all $30$ fractions,
for the FV plan (top row) as well as a 7-beam FI plan (middle row) and a 13-beam FI plan (bottom row).
Corresponding dose-volume histograms are shown in figure~\ref{hnk_dvh}.
Compared with the 7-beam FI plan, the FV plan achieves dosimetric improvements
for the cochleas (bilateral), chiasm, brainstem, parotids, and orbits.
Mean dose was reduced by 10.2 Gy (cochleas), 3.4 Gy (chiasm),
1.2 Gy (brainstem), and 1.0 Gy (orbits).
Max dose was reduced by 10.0 Gy (cochleas), 6.7 Gy (orbits), 5.7 Gy (brainstem),
4.7 Gy (parotids), 1.6 Gy (brain), and 1.4 Gy (pharynx).
The dosimetric quality of the FV plan is comparable to that of the 13-beam FI plan.
 
%\FloatBarrier
\subsection{Plan quality metrics} 
%\begin{table}[htbp]
\begin{table}[!htb]
\begin{center}
%\begin{ruledtabular}
\begin{tabular}{l l l l l l l}
\toprule
Case & $D_{95}$ (Gy) & $D_{98}$ (Gy) & $D_{99}$ (Gy) & $\text{D}_{\max}$& HI   \\ \midrule
PRT & 40.0 \color{blue} [40.0] & 39.8 \color{blue} [39.9] &39.6 \color{blue} [39.8] &40.9 \color{blue} [42.6] &.98 \color{blue} [.95] \\
LNG & 48.0 \color{blue} [48.0] & 47.6  \color{blue} [47.8] & 47.1 \color{blue} [47.7] & 49.6 \color{blue} [51.23] &.97 \color{blue} [.94]  \\
H\&N  & 66.0 \color{blue}[66.0] & 65.3 \color{blue} [65.6] & 64.5 \color{blue} [65.2] & 67.9 \color{blue} [68.3]& .97 \color{blue} [.97] \\
\bottomrule
\end{tabular}
%\end{ruledtabular}
\end{center}
\caption{PTV coverage metrics for several cases.  FV BOO results are shown in black, 
FI plan results in blue.  The homogeneity index HI is defined as $\text{D}95/\text{D}5$.}
\label{metrics}
\end{table}

%\begin{table}[htbp]
\begin{table}[!htb]
\begin{center}
%\begin{ruledtabular}
\begin{tabular}{l l l l l l l}
\toprule
&&\multicolumn{2}{c}{$\text{D}_{\text{mean}}^{\text{FV}} - \text{D}_{\text{mean}}^{\text{FI}}$}& &\multicolumn{2}{c}{$\text{D}_{\max}^{\text{FV}} - \text{D}_{\max}^{\text{FI}}$}\\
\midrule
Case &\quad & average (Gy) & range (Gy)& \qquad \qquad &  average (Gy) & range (Gy) \\ \midrule
PRT && -3.0 &$[-5.7,.24]$ && -1.8 &$[-6.0,.49]$\\
LNG && -.27 &$[-1.9,.47]$ && -.25 &$[-6.1,2.7]$\\
H\&N && -1.5 & $[-10.2,2.3]$& & -3.4 & $[-10.0,6.2]$\\
\bottomrule
\end{tabular}
%\end{ruledtabular}
\end{center}
\caption{OAR dose differences for several cases. 
For each case, the difference in mean dose
$\text{D}_{\text{mean}}^{\text{FV}} - \text{D}_{\text{mean}}^{\text{FI}}$
is computed for all OARs.
The min, max, and average differences in mean dose are listed in columns 2 and 3.
Likewise, the min, max, and average values of
$\text{D}_{\text{max}}^{\text{FV}} - \text{D}_{\text{max}}^{\text{FI}}$
are listed in columns 4 and 5.
}
\label{oarMetrics}
\end{table}

Tables \ref{metrics} and \ref{oarMetrics} show treatment plan quality metrics for the three cases
listed in table~\ref{patientTable1}.
For each case, metrics are reported for the FV plan as well as for the corresponding FI plan
that uses approximately the same number of beams per fraction as the FV plan.
(So for cases ``LNG'' and ``H\&N'' the FV plan is compared against the 7-beam FI plan, 
and for case ``PRT'' the FV plan is compared against the 10-beam FI plan.)
The FV plans show improvement in PTV $\text{D}_{\text{max}}$ in all cases,
as well as improvements in PTV homogeneity (defined as $D_{95}$/$D_5$) for cases ``PRT'' and ``LNG''.
There is a slight decrease in PTV $D_{98}$ and $D_{99}$.

For a given OAR, let $\text{D}_{\text{mean}}^{\text{FV}}$ and $\text{D}_{\text{mean}}^{\text{FI}}$ denote the mean doses delivered to the OAR
by the FV plan and FI plans, respectively.
Let $\text{D}_{\text{max}}^{\text{FV}}$ and $\text{D}_{\text{max}}^{\text{FI}}$ denote the maximum doses delivered to the
OAR by the respective plans.
For each case, we computed the difference in mean dose (that is, $\text{D}_{\text{mean}}^{\text{FV}} - \text{D}_{\text{mean}}^{\text{FI}}$)
and the difference in max dose (that is, $\text{D}_{\text{max}}^{\text{FV}} - \text{D}_{\text{max}}^{\text{FI}}$) for each OAR.
The results are summarized in table~\ref{oarMetrics}.

\section{Discussion}

%By allowing beam angles to vary between fractions, FV BOO exploits 
%degrees of freedom in IMRT which have traditionally been underutilized.
%As demonstrated by the results in section III, FV plans are able to improve dosimetry without increasing the number of beams per fraction, or alternatively to reduce the number of beams per fraction without compromising dosimetry.
%The FV~BOO strategy addresses a main criticism of non-coplanar IMRT treatment plans,
%which is the long delivery time due to time-consuming gantry and couch maneuvering for the large number of static beams.
%FV~BOO offers the potential to reduce this delivery time by using fewer beams per fraction.

By allowing beam angles to vary between fractions, FV BOO exploits 
degrees of freedom in IMRT which have traditionally been underutilized.
As demonstrated by the results in section III, FV plans are able to improve dosimetry without increasing the number of beams per fraction, 
or alternatively to reduce the number of beams per fraction without compromising dosimetry.
The FV~BOO strategy addresses a main criticism of non-coplanar IMRT treatment plans,
%which is the long delivery time due to time-consuming gantry and couch maneuvering for the large number of static beams.
which is the long delivery time due to the large number of static beams required
to sufficiently sample the vast non-coplanar beam angle space.
FV~BOO offers the potential to reduce this delivery time by using fewer beams per fraction,
while still utilizing a large \emph{total} number of non-coplanar beam angles.
An excellent example is provided by the improvement in dose compactness
observed in case ``LNG'' in section~\ref{numericalResults}.
It was shown in ~\cite{dong20134pi_2} that for lung SBRT cases non-coplanar IMRT
yields a substantial improvement in dose compactness as measured by $\text{R}_{50}$,
%but it was necessary to utilize well more than 10 non-coplanar beam angles to achieve this improvement.
but to achieve this improvement it was necessary to utilize well more than 10 non-coplanar beam angles.
In the lung SBRT case ``LNG'' studied here, the FV approach was able to achieve
superior dose compactness while using half as many beams per fraction as the FI plan.
Therefore, the FV planning method removes a major roadblock to implementing
$4\pi$ non-coplanar radiotherapy.

This study is novel in optimizing all beams for the entire plan simultaneously yet allowing different
sets of beam angles to be used in different fractions.
Aside from this central point, a key detail which distinguishes our approach is that the PTV is covered homogeneously in each fraction.
In contrast, the spatiotemporally non-uniform  fractionation schemes studied in~\cite{unkelbach2013simultaneous,unkelbach2015non,unkelbach2015emergence}
target subregions of the PTV at each fraction.  
While that is an effective strategy for BED-based optimization,
questions remain about the robustness of the resulting plans against setup error, 
and further work is needed to validate the biological effect of the heterogeneous PTV doses on tumor growth.
%Thus, the FV BOO approach presented here is potentially more clinically translatable. 
Thus, the FV BOO approach which maintains a uniform tumor dose coverage is potentially more clinically translatable. 
%The current method bears a conceptual similarity to SFUD planning.
%However, its ability to optimize all candidate beams together clearly distinguishes the two methods. 
FV~BOO bears a conceptual similarity to SFUD planning.
However, the ability of FV~BOO to optimize all candidate beams together clearly distinguishes the two methods. 
%Although the current method bears a conceptual similarity to SFUD planning,
%its ability to optimize all candidate beams together clearly distinguishes the two methods. 

This work is only a first step towards fraction-variant beam orientation
optimization, and there is likely room for improvement in both the
problem formulation and the optimization algorithm.
For example, if the per fraction dose to certain OARs poses a concern, the objective function can be modified to enforce fraction-wise OAR control: 
\begin{align}
\label{fracVariantProb2}
\mmz_x & \quad \sum_{f=1}^F \left( \underbrace{\frac12 \| A_0 x_f - d_0/F  \|_2^2
+ \sum_{i=1}^N \frac{\alpha_i}{2} \| (A_i x_f - d_i)_+ \|_2^2}_{\text{controls fractional doses to PTV and OARs}} \right)  \\
\notag & \quad + \underbrace{\sum_{i=1}^N \frac{\beta_i}{2}\| \bar A_i x \|_2^2}_{\text{controls total dose to OARs}} + \underbrace{\gamma \| Dx \|_1^{(\mu)}}_{\text{fluence map deliverability}}
+ \underbrace{\sum_{f=1}^F \sum_{b=1}^B w_b \| x_{f,b} \|_2^{1/2}}_{\text{group sparsity}} \\
\notag \subjto & \quad x \geq 0.
\end{align}

An additional area for improvement is the optimization runtime,
which in our experiments is about 60-90 minutes for non-coplanar five fraction plans
and about 7 hours for non-coplanar 30 fraction plans (with 500-700 candidate beams per fraction).
The computational expense for each FISTA iteration
is dominated by matrix-vector multiplications involving the large sparse matrices $A_i$
and $A_i^T$. These matrix-vector multiplications could be made much faster with a GPU implementation.
Besides improving the implementation, it may be possible to improve
the algorithm as well. 
While we have found FISTA to be effective for this application, other algorithms should be investigated,
such as truncated interior point methods~\cite{kim2007interior,koh2007interior,portugal2000truncated},
or recently developed proximal quasi-Newton methods~\cite{becker2012quasi,scheinberg2016practical,lee2014proximal}. 
New algorithmic innovations such as performing column clustering on the dose-calculation matrix to reduce
the problem size,
as suggested in~\cite{ungun2017real}, will likely be beneficial.
Note that we have found that if the OAR weights $\alpha_i$ and $\beta_i$
are tuned for the case where there is only one treatment fraction,
then the same weights tend to yield a good selection of beams for
the full fraction-variant problem with $F$ fractions.
This is very helpful because in our experiments it only takes about 5-8 minutes to solve
problem~\eqref{fracVariantProb} when $F = 1$.

One potential burden of fraction-variant BOO is increased patient-specific IMRT quality assurance (QA) load. 
It would be time-consuming to QA all individual fractions. 
Measurement of individual beams may be feasible but may lead to difficulty in analyzing the results 
because of the non-uniform target dose resulting from a single beam. 
However, this is unlikely to be an insurmountable difficulty as machine log file analysis in combination 
with independent dose calculation has shown efficacy equivalent to measurement-based QA~\cite{sun2012evaluation}.

\section{Conclusions} 
\label{conclusions}

This work demonstrates the first beam orientation optimization algorithm that simultaneously optimizes beam angles
for all IMRT treatment fractions.
The resulting fraction-variant plans offer improved dosimetry without increasing the beam budget
(that is, the number of beams per fraction that are utilized).
Alternatively, fraction-variant plans allow the beam budget to be reduced without
compromising dosimetry.

%In the cases that we investigated, the dosimetric quality of the fraction-variant plans is superior 
%to that of conventional plans that use approximately the same
%number of beams per fraction.  Moreover, the fraction-variant plans are dosimetrically similar to conventional plans
%that use twice as many beams per fraction.

\section*{Acknowledgements}
This research was funded by NIH grants R43CA183390, R44CA183399, R01CA188300,
and Department of Energy grants DE-SC0017057 and DE-SC0017687.

\section*{Disclosure of conflicts of interest}
The authors have no relevant conflicts of interest to disclose.

\appendix

\section{Prox-operator calculation}
\label{proxAppendix}
Here we derive a formula for the prox-operator of the function $h:\mathbb R^n \to \mathbb R \cup \{\infty\}$
defined by
\[
h(x) = \begin{cases} \|x\|_2^{p} & \quad \text{if } x \geq 0 \\
\infty & \quad \text{otherwise},
\end{cases}
\]
in the special case where $p = 1/2$.
(The inequality $x \geq 0$ is interpreted componentwise.)  
Let $t > 0$. To evaluate $\prox_{th}(\hat x)$,
we must find the minimizer for the optimization problem
\begin{align}
\label{proxProb}
\mmz_x & \quad \|x\|_2^p + \frac{1}{2t} \|x - \hat x \|_2^2 \\
\notag \subjto & \quad x \geq 0.
\end{align}
First note that if $\hat x_i \leq 0$ then there is no benefit
from taking the component $x_i$ to be positive.  If $x_i$ were positive,
then both terms in the objective function could be reduced just by setting
$x_i = 0$.

It remains only to select values for the other components of $x$.
This is a smaller optimization problem, with one unknown for each positive
component of $\hat x$.  The negative components of $\hat x$ are irrelevant
to the solution of this reduced problem.  Thus, we would still arrive at the same
final answer if the negative components of $\hat x$ were set equal to $0$
at the very beginning.

In other words, problem \eqref{proxProb} is equivalent to the problem
\begin{align*}
\mmz_x & \quad \|x\|_2^p + \frac{1}{2t} \|x - \max(\hat x,0) \|_2^2 \\
\notag \subjto & \quad x \geq 0,
\end{align*}
which in turn is equivalent to the problem
\begin{align*}
\mmz_x & \quad \|x\|_2^p + \frac{1}{2t} \|x - \max(\hat x,0) \|_2^2 \\
\end{align*}
(because there would be no benefit from taking any components of $x$
to be negative).  This shows that
\begin{equation}
\label{proxFormula}
\prox_{th}(\hat x) = \prox_{t \| \cdot \|_2^p}(\max(\hat x,0)).
\end{equation}

Formula~\eqref{proxFormula} is valid for any $p > 0$.
The reason we take $p = 1/2$ is that a short and explicit 
(but non-obvious) formula
for the prox-operator of the function $\psi(y) = \| y \|_2^{1/2}$ is available
\cite{mollenhoff2015low}.
To evaluate $\prox_{t \psi}(y)$, first let
$\alpha = t / \|y\|_2^{3/2}$. 
(If $y = 0$ then $\alpha = \infty$.) Then
\begin{equation}
\label{proxL2OneHalf}
\prox_{t \psi}(y) = s^2 y, \qquad \text{where} \quad
s =
\begin{cases}
\frac{2}{\sqrt{3}} \sin(\frac13 \arccos(\frac{3 \sqrt{3}}{4} \alpha) + \frac{\pi}{2}) & \
\quad \text{if } \alpha \leq \frac{2\sqrt{6}}{9}, \\
0 & \quad \text{otherwise.}
\end{cases}
\end{equation}

\section{The choice of group sparsity exponent $p$.}
\label{pAppendix}
Here we explain why the choice $p = 1$ for the exponent in the group sparsity penalty function is forbidden.
Suppose we take $p = 1$, so that problem~\eqref{fracVariantProb} is convex,
and let $x^\star$ be a minimizer for~\eqref{fracVariantProb}.
The blocks of $x^\star$ are denoted $x^\star_1,\ldots, x^\star_F$.
We can obtain another minimizer for problem~\eqref{fracVariantProb}
by permuting the blocks of $x^\star$.  In other words, the
vector $x^\star_\sigma$ defined by
\[
x^\star_\sigma = \begin{bmatrix} x^\star_{\sigma(1)} \\ x^\star_{\sigma(2)} \\ \vdots \\ x^\star_{\sigma(F)} \end{bmatrix}
\]
is optimal for~\eqref{fracVariantProb} for any permutation~$\sigma$
of $\{1,2,\ldots,F\}$.
In fact, if $S_F$ is the set of all permutations of $\{1,\ldots,F\}$, then
any convex combination of vectors $x^\star_\sigma$ (with $\sigma \in S_F$) is optimal as well:
\[
x_\theta = \sum_{\sigma \in S_F} \theta_\sigma x^\star_\sigma
\]
is optimal for problem~\eqref{fracVariantProb}
whenever $\sum_{ \sigma \in S_F} \theta_\sigma = 1$
and $\theta_\sigma \geq 0$ for all $\sigma \in S_F$.
An optimization algorithm that finds a global minimizer for problem~\eqref{fracVariantProb}
(with $p = 1$) will almost surely find one of the solutions
$x_\theta$ with $\theta_\sigma > 0$ for all $\sigma \in S_F$,
and such a solution has the same set of active beams for each fraction.  
We have observed this phenomenon in numerical experiments.
Thus, FV BOO requires the use of a non-convex group sparsity penalty term.
%Thus, the function $\omega_1$
%fails as a surrogate for $\omega_0$,
%and we are forced to use a non-convex group sparsity penalty term.
%(Also note that the group restricted isometry property is not satisfied
%for problem~\eqref{fracVariantProb} because the matrices $\bar A_i$
%have repeated block columns.)
We choose $p = 1/2$ specifically because the proximal operator of the function
$\psi(x) = \| x \|_2^{1/2}$ can be evaluated easily
using formula~\eqref{proxL2OneHalf} in appendix~\ref{proxAppendix}.

\section{Selecting the weights $w_b$ in the group sparsity term.}
\label{weightAppendix}
In this section we explain our method for choosing the weights
$w_b$ that appear in the group sparsity term in problem~\eqref{fracVariantProb}.
Some beams must only travel a short distance through the body to reach the PTV,
whereas other ``long path'' beams must travel a greater distance through the body
before reaching the PTV.  
To overcome attenuation, a ``long path'' beam must be fired more intensely
than a short path beam in order to deliver the same dose to the PTV.
If all the weights $w_b$ in the group sparsity term are chosen to be equal, 
then the group sparsity penalty
introduces a bias in favor of short path beams, because a long path
beam $b$ requires $\| x_b \|_2$ to be large in order to target the PTV effectively.
We choose the weights $w_b$ to compensate for this bias.

Let $n_b$ be the number of beamlets in beam $b$ with a trajectory that intersects the PTV.
Suppose that beam $b$ is fired uniformly, so that $x_b = \lambda \vec 1$,
and the scalar $\lambda$ is chosen so that the mean dose delivered to the PTV
by beam $b$ is $1$ Gy.  Then it is easy to check that 
$\| x_b \|_2^p = \left(\sqrt{n_b}/\text{mean}(A_0^b \vec 1 \,)\right)^p$,
where $A_0^b$ is the dose-calculation matrix from beam $b$
to the PTV.  This quantity is larger for long path beams than for short path beams.
Therefore, to level the playing field, we choose the weights $w_b$
so that
\begin{equation}
\label{weightFormula}
w_b = c \left(\frac{ \text{mean}(A_0^b \vec 1 \,)}{\sqrt{n_b}} \right)^p. % PICK UP HERE, FIX THIS FORMULA.  FINISH INTRO. FIX ABSTRACT.
\end{equation}
The scalar $c$ is chosen to be the same for all beams,
and $c$ is tuned by trial and error to achieve the desired
group sparsity level.

%\bibliographystyle{unsrt}
%\bibliography{D:/Users/Documents/GitHub/fistaPaper/myBibFile}

%\printbibliography

\end{document}